\documentclass[journal]{IEEEtran}

\usepackage{enumitem}
\usepackage{amsmath,graphics,amssymb,epsfig,subfigure,color,cite}
\usepackage{array}
\usepackage{multirow}
\usepackage{enumerate}
\usepackage{algorithm2e}
\usepackage{algpseudocode}
\usepackage{algorithmicx}
\usepackage{algcompatible}
\usepackage{graphics}

\usepackage{bm}
\usepackage{float}
\usepackage{makecell}
\usepackage[right=0.625in, left=0.625in, top=0.75in, bottom=1.0in]{geometry}

\newcommand{\argmin}{\operatornamewithlimits{argmin}}
\makeatletter
\newcommand{\vast}{\bBigg@{4.5}}
\newcommand{\Vast}{\bBigg@{7.5}}
\makeatother

\begin{document}

    \title{Deep Learning-Based CSI Feedback for Wi-Fi Systems With Temporal Correlation}

    \author{Junyong Shin, Eunsung Jeon, Inhyoung Kim, and Yo-Seb Jeon 
    	\thanks{Junyong Shin and Yo-Seb Jeon are with the Department of Electrical Engineering, POSTECH, Pohang, Gyeongbuk, Republic of Korea  (e-mail: sjyong@postech.ac.kr, yoseb.jeon@postech.ac.kr).}
        \thanks{Eunsung Jeon and Inhyoung Kim are with Connectivity Development Team, Samsung Electronics Co., Ltd., Republic of Korea  (e-mail: eunsung.jeon@samsung.com, inhyoung.kim@samsung.com).}
    }
	\vspace{-2mm}
	
	\maketitle
    \thispagestyle{empty}
  \pagestyle{empty}
	\vspace{-12mm}

	\begin{abstract} 
        To support higher throughput in next-generation Wi-Fi systems, efficient compression and feedback of channel state information (CSI) from a station (STA) to an access point (AP) is essential. This paper proposes a deep learning (DL)-based CSI feedback framework tailored for Wi-Fi systems. The framework employs encoder and decoder networks to compress and reconstruct CSI angle parameters, with a trainable vector quantization (VQ) module enabling efficient finite-bit representation through end-to-end training. To further enhance performance, we introduce an angle-difference feedback strategy that exploits the temporal correlation of the angle parameters by feeding back the difference between the current and previous values. This is complemented by preprocessing that handles the periodicity of angles and tailored VQ modules that compensate for residual quantization errors. Additionally, we present a DL-based CSI refinement module at the AP, which improves reconstruction by jointly using current and prior feedback. Simulation results show that the proposed framework outperforms both standard Wi-Fi feedback  and existing DL-based feedback methods, with notable gains from both angle-difference feedback and CSI refinement.
	\end{abstract}
	

	\section{Introduction}\label{Sec:Intro}
\begin{figure}[t]
        \centering 
        {\epsfig{file=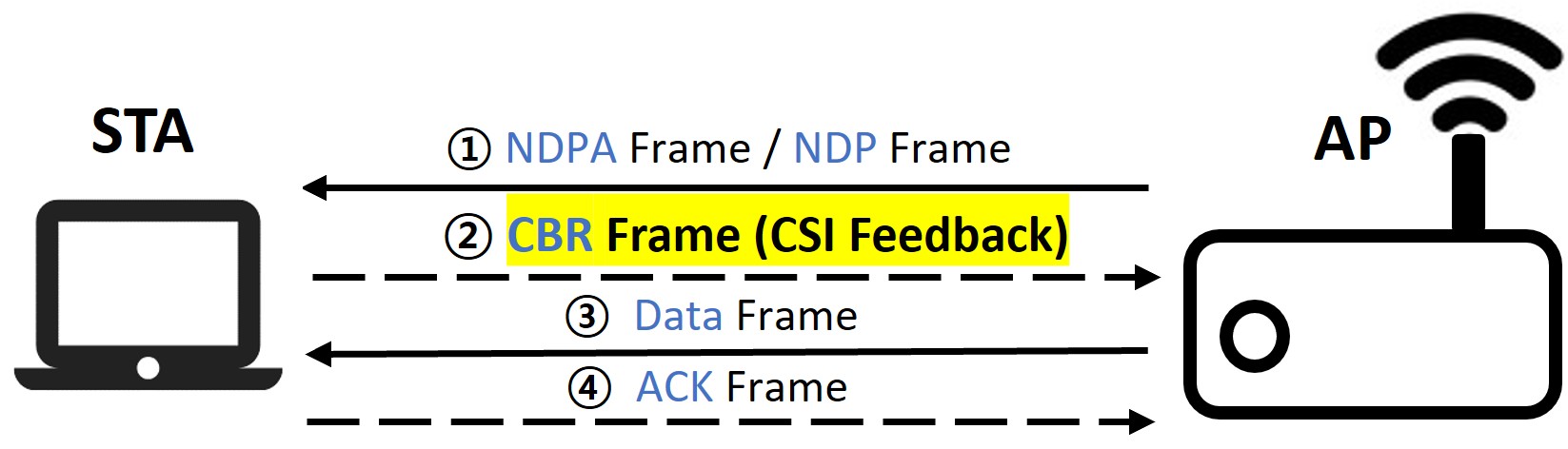, width=7.5cm}}
    
        \caption{An illustration of a communication protocol in Wi-Fi systems.}
        \label{fig: protocol}\vspace{-3mm}
    \end{figure}

    Recently, wireless networks have dramatically advanced to meet the growing demand for higher data rates. Specifically, the IEEE 802.11 task group has been developing a novel extremely high throughput (EHT) amendment for wireless local area networks (WLANs) \cite{Wi-Fi7}. With developing an advanced multiple-input multiple-output (MIMO) techniques, it aims to service much higher throughput and reliable communications.
    To achieve higher throughput in next-generation Wi-Fi systems utilizing MIMO techniques, acquiring an accurate downlink channel state information (CSI) is indispensable for the access point (AP) to transmit data to the station (STA). The typical CSI acquisition process in the system is initiated by the AP's transmitting the null data packet announcement (NDPA) frame to STA. NDPA serves the role of notifying the STA of all necessary probing details of MIMO transmission. Then, a null data packet (NDP) is transmitted to the STA for estimating the downlink CSI. After the channel estimation, the STA compresses and feeds the CSI back to the AP, in the form of the compressed beamforming report (CBR). The compression is performed by extracting angle parameters after performing singular value decomposition (SVD) of the MIMO channel\cite{Wi-Fi5,Wi-Fi6,Wi-Fi7}. The aforementioned protocol in the Wi-Fi systems is summarized in Fig.~\ref{fig: protocol}. Another approach to acquire the downlink CSI is to leverage channel reciprocity. In this process, the downlink CSI is inferred from the uplink CSI with additional calibrations. However, this approach is excluded from IEEE 802.11ac standard \cite{Wi-Fi5}, indicating its verified limitation.

    The CSI feedback methodologies have been an important problem for massive MIMO systems in cellular networks as well. Among various methodologies, deep learning (DL)-based CSI feedback methods have been proposed for the cellular networks in recent years \cite{Overview, CsiNet, CsiNet+, CRNet,ReNet, DCRNet}.
    By exploiting the sparsity of CSI in massive MIMO systems and carefully designing neural networks, these methods have demonstrated their effectiveness in reducing CSI feedback overhead, outperforming traditional non-DL-based approaches like compressed sensing and static codebook-based feedback\cite{conventional1, conventional2}.
    A typical approach of the DL-based CSI feedback is to compress a preprocessed CSI into a low-dimensional latent vector by using an encoder network. Then, the CSI is reconstructed by using a decoder network from a received latent vector.
    To enable the transmission of the latent vector with a finite number of bits, one possible approach is to apply scalar quantization (SQ) to each latent element. For example, non-uniform SQ methods were explored in \cite{CQNet, PQB} to represent the latent vector with a limited number of bits for CSI feedback in cellular networks. However, the application of SQ to DL-based CSI feedback has inherent limitations. First, SQ disregards the correlations between latent elements, resulting in suboptimal performance when these elements are correlated \cite{VQpros.}. Second, SQ cannot allocate fewer than one quantization bit per latent entry, which imposes constraints on the dimensionality of the latent vector for a given feedback overhead and ultimately limits the feedback efficiency.
    These limitations can be resolved by applying a vector quantization (VQ) in the latent space.
    The application of VQ to the latent representation allows for modeling the correlation among latent entries. Furthermore, the joint optimization of the VQ codebook with the entire model facilitates quantization that effectively reflects the distribution of the latent vectors, thereby enhancing the efficiency of the learned representations \cite{VQ-VAE}.
    Motivated by this, DL-based CSI feedback methods incorporating VQ have been proposed for cellular networks in \cite{shape-gain, ICTC, ECVQ}. In \cite{shape-gain}, it was reported that the DL-based CSI feedback with the VQ framework can achieve a better CSI reconstruction performance than that of DL-based CSI feedback with the SQ framework in cellular networks.

    To further improve the efficiency of DL-based CSI feedback in massive MIMO systems, some studies have leveraged the temporal correlation of CSI, which naturally arises in practical communication scenarios involving devices with low to moderate mobility \cite{CsiNet-LSTM, LSTM_CSI,Differential,Markovian, Comment3.4_1,Comment3.4_2,Comment3.4_3}. A common approach in this direction is to employ neural network architectures specifically designed for time-series inputs. For instance, it was reported in \cite{CsiNet-LSTM, LSTM_CSI} that incorporating long short-term memory (LSTM) \cite{LSTM} enables the model to capture temporal correlation, enhancing feedback performance. On the other hand, an alternative approach to leveraging the temporal correlation of CSI is to exploit the inherent sparsity in the differences between sequential CSI instances. This approach stems from the intuition that sequential instances are likely to exhibit similar values, leading to sparse differences that are more efficient to compress and reconstruct. The sparsity in CSI differences was leveraged in \cite{Differential,Markovian}, demonstrating an efficacy of this approach in improving the performance of the CSI feedback in cellular networks. 
    

    
    Unlike DL-based CSI feedback in cellular networks, limited attention has been given to DL-based CSI feedback for Wi-Fi systems \cite{LB-SciFi,EFNet}. In fact, existing DL-based CSI feedback methods for cellular systems are developed under massive MIMO assumptions and rely on sparsity of CSI in the angular-delay domain, which does not hold in typical Wi-Fi systems. Furthermore, these methods do not align with current Wi-Fi system protocols, making their direct application much more impractical. From these points, in \cite{LB-SciFi}, the angle parameters of downlink CSI in Wi-Fi systems were separately compressed using hyperparameterized preprocessing and two distinct fully connected networks (FCNs). In \cite{EFNet}, a convolutional neural network (CNN)-based model with a channel attention module was used to compress a right singular matrix in SVD, rather than the angle parameters. However, these studies employ simple quantization techniques without jointly training the quantization module with the encoder and decoder networks. Moreover, none of the existing research has leveraged the temporal correlation of CSIs in DL-based CSI feedback specifically for Wi-Fi systems. The potential benefits of the joint training of the quantization module and the utilization of temporal correlation in CSIs remain unexplored, despite their critical importance in realizing the full potential of DL-based CSI feedback in next-generation Wi-Fi systems.


    To fill this research gap, in this paper, we propose novel DL-based CSI feedback and refinement frameworks that pioneer the use of a trainable VQ module and the exploitation of temporal correlation in CSIs for Wi-Fi systems. The major contributions in this paper are summarized as follows:
    \begin{itemize}


    \item We present a novel DL-based CSI feedback framework tailored for Wi-Fi systems, which incorporates both initial and angle-difference feedback strategies. For the initial feedback, we determine an appropriate input representation by aggregating angle parameters defined in the IEEE 802.11 standard and employ a trainable VQ module jointly optimized with the encoder and decoder to enable efficient finite-bit compression. For the angle-difference feedback, we exploit the temporal correlation of CSI by feeding back the difference between consecutive angle parameters. To ensure accurate encoding of these differences, we introduce a preprocessing step that accounts for the periodicity of angles, and design two VQ architectures—{\em parallel} VQ and {\em unified} VQ—to mitigate error accumulation from previous feedback steps.


    \item We propose a new CSI refinement framework for Wi-Fi systems that enhances reconstruction performance at the AP by leveraging the temporal correlation of CSIs.
    We refine angle parameters using both current and previous feedback, and its performance is further improved by a recursive strategy that feeds refined outputs back into future refinements. This framework enables more accurate and efficient CSI reconstruction on the AP side.


    \item Using extensive simulations, we validate the effectiveness of the proposed CSI feedback and refinement frameworks for Wi-Fi systems with temporally correlated channels. The results show that the proposed frameworks consistently deliver significant performance gains across various scenarios. Moreover, the proposed frameworks outperform the standard feedback method used in current Wi-Fi systems in terms of both CSI reconstruction accuracy and net throughput.
    \end{itemize}

    This work builds upon our previous study \cite{Junyong_Globecom}, where we introduced only the DL-based angle-difference CSI feedback framework. In the current work, we newly propose a DL-based CSI refinement strategy to further improve the feedback efficiency by leveraging the temporal correlation of CSIs on the AP side.
    To refine the angle parameters effectively at each time step, we design a CSI refinement module and a recursive strategy that utilize both previous and current feedback.
    With this extension, we enhance the simulation study to demonstrate the superiority of our framework in Wi-Fi systems. Our simulation results demonstrate the superiority of the proposed CSI refinement framework in enhancing the reconstruction performance and efficiency of the CSI feedback in Wi-Fi systems.

    \vspace{2mm}
    {\em Notations:} Throughout this paper, matrices, vectors, and scalars are denoted by bold uppercase, bold lowercase, and normal lowercase letters, respectively.
     The superscripts $(\cdot)^{\rm T}$, $(\cdot)^{*}$, and $(\cdot)^{\rm H}$  denote the transpose, conjugate, and hermitian of a vector/matrix, respectively. Also, Re\{$\cdot$\} and phase($\cdot$) are the operators which return the real component and phase component of an input. Furthermore, diag($\cdot$) denotes a diagonal matrix whose diagonal components consist of an input vector. For the other operators, $\mathbb{E}\{\cdot\}$ , $|\cdot|$, $\Vert\cdot\Vert$, and $\Vert\cdot\Vert_{\rm F}$ represent the expectation operator, absolute value, 2-norm of a vector, and Frobenius norm of a matrix, respectively.
     We refer to an element of a (possibly multidimensional) matrix $\mathbf{\Phi}$ at position $(l, m, n)$ as $[\mathbf{\Phi}]_{l,m,n}$, where the number of indices depends on the dimension of $\mathbf{\Phi}$.
     When referring to a group of values in a certain range of the sequence/matrix, a colon is used as $\mathbf{\Phi}_{t:t+T}$ or $[\mathbf{V}]_{i:N,j}$. Furthermore, $\mathbb{R}$ and $\mathbb{C}$ represent the real and complex numbers, respectively.

	
    \section{System Model and preliminary}\label{Sec:Model}
    \subsection{Downlink MIMO Wi-Fi Systems}
    Consider a downlink Wi-Fi system in which an AP equipped with $N_t$ transmit antennas transmits data to a STA equipped with $N_r$ receive antennas. The system employs orthogonal frequency division multiplexing (OFDM) with $N_c$ subcarriers. To facilitate channel estimation at the STA, the AP transmits the NDP to the STA and a received NDP can be expressed as
    \begin{align}
        \mathbf{y}[k] = \mathbf{H}[k]\mathbf{s}[k] + \mathbf{n}[k],\quad k=0,1,...,N_c-1,
    \end{align}
    where $\mathbf{y}[k]\in \mathbb{C}^{N_r}, \mathbf{H}[k]\in \mathbb{C}^{N_r\times N_t}, \mathbf{s}[k] \in \mathbb{C}^{N_t},$ and $\mathbf{n}[k] \in \mathbb{C}^{N_r}$ represent the received NDP, channel matrix, transmitted NDP, and additive noise vector respectively. Then, the STA estimates the CSI (i.e., channel frequency responses (CFRs) associated with all subcarriers) using the received NDP. The channel estimation is assumed to be perfect in this work. In the IEEE 802.11 standard protocols, to extract beamforming matrices, the STA computes the SVD of each CFR as
    \begin{align}
    \mathbf{H}[k]=\mathbf{U}[k]\mathbf{\Sigma}[k]\tilde{\mathbf{V}}^{\text H}[k],
    \end{align}
    where $\mathbf{U}[k]\in\mathbb{C}^{N_r \times N_r}$ and $\mathbf{\tilde V}[k]\in\mathbb{C}^{N_t \times N_t}$ are unitary matrices, and $\mathbf{\Sigma}[k]\in\mathbb{C}^{N_r \times N_t}$ is a diagonal matrix. In the SVD beamforming, $\mathbf{U}[k]$ and $\mathbf{\tilde V}[k]$ are utilized to design receive and transmit beamforming matrix, respectively. To this end, in the IEEE 802.11 protocols, the first $N_s$ columns of $\tilde{\mathbf{V}}[k]$, denoted as $\overline{\mathbf{V}}[k]\in\mathbb{C}^{N_t\times N_s}$, are compressed and fed back to the AP, where $N_s$ represents the number of data streams.

    Now, we present the CSI compression method employed in the IEEE 802.11 standard, which is based on the {\em Givens rotation} with angle parameters $(\phi,\psi)$\cite{givens1,givens2,Kmeans_V}.
    Givens rotation is a technique that expresses a matrix with orthonormal columns as a product of successive plane rotations, and it has been utilized not only for CSI parameter extraction but also in a range of applications, including beamforming methodologies\cite{givens_beam1,givens_beam2}.
    First, the phase offset of the last row of $\overline{\mathbf{V}}[k]$ is eliminated as
    \begin{align}
        \mathbf{V}[k] = \mathbf{\overline V}[k]\mathbf{\tilde{D}}^*[k],
    \end{align}
    where $\mathbf{\tilde{D}}[k]\in\mathbb{C}^{N_t\times N_t}$ is a diagonal matrix, whose diagonal components are composed of phase coefficients of the last row of $\mathbf{\overline V}[k]$. Note that it is sufficient for the AP to use the transmit beamforming with $\mathbf{V}[k]$, since $\mathbf{\tilde{D}}[k]$ is a diagonal matrix and exchangable with $\mathbf{\Sigma}[k]$, so that it can be equalized by STA. This ensures that last row of $\mathbf{V}[k]$ consists of non-negative real values.

    \begin{figure*}[t]
        \centering 
        {\epsfig{file=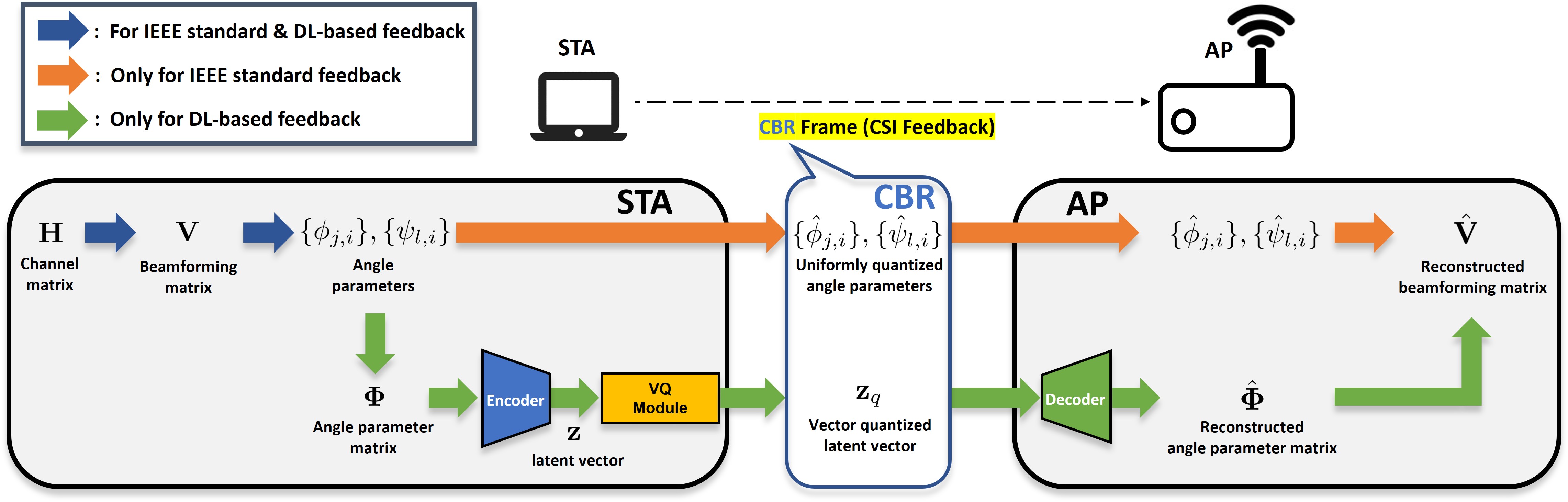, width=15cm}}
    
        \caption{An illustration of overall CSI feedback procedures for IEEE 802.11 standard and DL-based methods.}
        \label{fig: overallflow}\vspace{-3mm}
    \end{figure*}
    
    Next, $\mathbf{V}[k]$ is rewritten as the product of the Givens rotation matrix as
    \begin{align} \label{eq: angle_to_V}
    &\mathbf{V}[k] \nonumber \\
    &=\left(\prod^{{\rm min}(N_s,N_t-1)}_{i=1} \left(\mathbf{D}_i\big(\boldsymbol{\phi}_i^{(k)}\big)\prod^{N_t}_{l=i+1}\mathbf{G}_{li}^{\sf T}\big(\psi_{li}^{(k)}\big) \right) \right)\mathbf{I}_{N_t\times N_s},
    \end{align}
    where $\mathbf{D}_i\left(\boldsymbol{\phi}_i\right)$ is an $N_t\times N_t$ diagonal matrix defined as
    \begin{align}
    \mathbf{D}_i(\boldsymbol{\phi}_i)=
    \begin{bmatrix}
    \mathbf{I}_{i-1} & 0 & 0 & \cdots & 0\\
    0 & e^{j\phi_{i,i}} & 0 & \cdots & 0\\
    0 & 0 & \ddots & 0 & 0\\
    0 & 0 & \cdots & e^{j\phi_{N_t-1,i}} & 0\\
    0 & 0 & \cdots & 0 & 1
    \end{bmatrix},
    \end{align}
    and $\mathbf{G}_{li}\left(\psi_{li}\right)$ is an $N_t\times N_t$ Givens rotation matrix defined as
    \begin{align}
    \mathbf{G}_{li}(\psi_{li})=
    \begin{bmatrix}
    \mathbf{I}_{i-1} & 0 & 0 & 0 & 0\\
    0 & \text{cos}\psi_{li} & 0 & \text{sin}\psi_{li} & 0\\
    0 & 0 & \mathbf{I}_{l-i-1} & 0 & 0\\
    0 & -\text{sin}\psi_{li} & 0 & \text{cos}\psi_{li} & 0\\
    0 & 0 & 0 & 0 & \mathbf{I}_{N_t-l}
    \end{bmatrix}.
    \end{align}
    Also, $\mathbf{I}_m$ represents an $m\times m$ identity matrix and $\mathbf{I}_{m\times n}$ represents an $m\times n$ ($ m\geq n$) matrix which corresponds to zero-padded version of $\mathbf{I}_m$.
    In the case of $N_r=N_t=N_s=3$, for instance, $\mathbf{V}[k]$ can be decomposed as  
    \begin{align}
    \mathbf{V}[k]=&
    \begin{bmatrix}
    e^{j\phi^{(k)}_{11}} & 0 & 0\\
    0 & e^{j\phi^{(k)}_{21}} & 0\\
    0 & 0 & 1
    \end{bmatrix} \times
    \begin{bmatrix}
    \text{cos}\psi^{(k)}_{21} & \text{sin}\psi^{(k)}_{21} & 0\\
    -\text{sin}\psi^{(k)}_{21} & \text{cos}\psi^{(k)}_{21} & 0\\
    0 & 0 & 1
    \end{bmatrix}^{\rm T}\nonumber \\ \nonumber 
    &\times
    \begin{bmatrix}
    \text{cos}\psi^{(k)}_{31} & 0 & \text{sin}\psi^{(k)}_{31}\\
    0 & 1 & 0\\
    -\text{sin}\psi^{(k)}_{31} & 0 & \text{cos}\psi^{(k)}_{31}
    \end{bmatrix}^{\rm T}
    \times
    \begin{bmatrix}
    1 & 0 & 0\\
    0 & e^{j\phi^{(k)}_{22}} & 0\\
    0 & 0 & 1
    \end{bmatrix} \\  
    &\times
    \begin{bmatrix}
    1 & 0 & 0\\
    0 & \text{cos}\psi^{(k)}_{32} & \text{sin}\psi^{(k)}_{32}\\
    0 & -\text{sin}\psi^{(k)}_{32} & \text{cos}\psi^{(k)}_{32}\\
    \end{bmatrix}^{\rm T} \times
    \begin{bmatrix}
    1 & 0 & 0\\
    0 & 1 & 0\\
    0 & 0 & 1
    \end{bmatrix},
    \end{align}
    with angle parameters $[\phi^{(k)}_{11},\phi^{(k)}_{21},\psi^{(k)}_{21},\psi^{(k)}_{31},\phi^{(k)}_{22},\psi^{(k)}_{32}]^{\rm T}$ extracted for the $k$-th subcarrier. From \eqref{eq: angle_to_V}, the angle parameters $\{\phi_{j,i}\}$ and $\{\psi_{l,i}\}$ can be extracted using Algorithm \ref{alg: V_to_angle}.

    {\small \RestyleAlgo{ruled}
    \SetKwComment{Comment}{/* }{ */}
    \begin{algorithm}[t]\label{alg: V_to_angle}
    \caption{Algorithm for extracting $\phi$ and $\psi$ from $\mathbf{\tilde V}[k]$}
    \textbf{1. Parameters:}
    $\boldsymbol{\phi}^{(k)}_i=[\phi^{(k)}_{i,i},...,\phi^{(k)}_{N_t-1,i}]$, $\psi^{(k)}_{li}$\\
    \textbf{2. Extraction:}\\
    \For{$k=0,...,N_c-1$}{
        $[\mathbf{U},\mathbf{\Sigma},\mathbf{\tilde V}] \leftarrow \text{SVD}(\mathbf{H}[k])$;\\
        $\mathbf{\overline V}\leftarrow[\mathbf{\tilde V}]_{:,1:N_s}$;\\
        $\mathbf{\tilde D} \leftarrow {\rm diag}({\rm exp}(j\times{\rm phase}([\mathbf{\overline V}]_{N_t,:})))$;\\
        $\mathbf{V}\leftarrow \mathbf{\overline V}\mathbf{\tilde D}^* $;\\
        \For{$i=1,...,{\rm min}(N_s,N_t-1)$}{
            $\boldsymbol{\phi}^{(k)}_i \leftarrow {\rm phase}([\mathbf{V}]_{i:N_t-1,i})$;\\
            $\mathbf{V} \leftarrow \mathbf{D}_i(\boldsymbol{\phi}^{(k)}_i )^{*}\mathbf{V}$;\\
            \For{$l=i+1,...,N_t$}{
                $\psi^{(k)}_{li}={\rm tan}^{-1}\left(\frac{{\rm Re}\{[\mathbf{V}]_{l,i}\}}{{\rm Re}\{[\mathbf{V}]_{i,i}\}}\right)$;\\
                $\mathbf{V} \leftarrow \mathbf{G}_{li}(\psi^{(k)}_{li} )^{*}\mathbf{V}$;\\
            }
        }
    }
    \end{algorithm}}

     Last, the angle parameters are uniformly quantized using $b_{\phi}$ and $b_{\psi}$ bits, respectively. The quantized parameters are expressed as
    \begin{align}
        \hat{\phi}_{j,i} &\in\left\{ \frac{n\pi}{2^{b_{\phi}-1}} + \frac{\pi}{2^{b_{\phi}}}\ \big|\ n=1,2,..., 2^{b_{\phi}-1}\right\}, \\
        \hat{\psi}_{l,i} &\in\left\{ \frac{n\pi}{2^{b_{\psi}+1}} + \frac{\pi}{2^{b_{\psi}+2}}\ \big|\ n=1,2,..., 2^{b_{\psi}-1}\right\}.
    \end{align}
    These parameters are fed back to the AP in order of $[...,\ \hat{\phi}^{(k)}_{i:N_t-1,i},\ \hat{\psi}^{(k)}_{i+1:N_t,i},\ ...]^{\rm T}$ for the $k$-th subcarrier.
    The aforementioned feedback method provides satisfactory performance when the number of feedback bits is sufficient, which is why it has been adopted in the IEEE 802.11 standard to date. However, when feedback overhead is limited, this method suffers from significant performance degradation, potentially creating a bottleneck in next-generation Wi-Fi systems. This limitation naturally motivates the development of a new CSI feedback method based on DL. The aforementioned IEEE standard feedback procedure is illustrated in Fig.~\ref{fig: overallflow}, alongside a DL-based alternative, which will be discussed in the following sections.

    \subsection{DL-based CSI Feedback Process}
    To implement a typical DL-based CSI feedback method for Wi-Fi systems, the STA employs an encoder network to compress the CSI input into the form of a latent vector with dimension $M$. 
    Meanwhile, the AP employs a decoder network to reconstruct the original CSI from the STA's feedback.

    The STA initiates a CSI feedback process by utilizing the CSI input $\mathbf{X}$ for the encoder network. Note that our choice for this input will be discussed in Sec.~\ref{Sec: VQ-VAE}.
    This yields the latent vector ${\bf z}$ expressed as ${\bf z} = f_{\text{enc}}(\mathbf{X})$, where $f_{\text{enc}}$ represents the encoder network. 
    Typically, the latent vector ${\bf z}$ is considered as a STA's feedback that needs to be transmitted to the AP.
    Upon the reception of the feedback, the AP reconstructs the CSI by utilizing the latent vector as an input of the decoder network $f_{\text{dec}}$. In practice, the latent vector ${\bf z}$ is transformed into a finite-length bit sequence before being transmitted to the AP. This transformation is achieved through quantizing the latent vector $\mathbf z$ under the constraint on feedback overhead. Subsequently, the AP reconstructs the CSI by utilizing the quantized latent vector, denoted as ${\bf z}_q$, as an input for the decoder network, i.e., $\mathbf{\hat X} = f_{\text{dec}}(\mathbf{z}_q)$. 

    \section{Proposed DL-based CSI Feedback Framework}\label{Sec: AD}
    In this section, we propose a novel DL-based CSI feedback framework which utilizes VQ approach to realize the efficient finite-bit feedback of angle parameters in Wi-Fi systems, while leveraging the temporal correlation of these parameters.
    First, we introduce an initial CSI feedback method that employs VQ to efficiently compress and feedback individually obtained CSI. Then, we present an angle-difference feedback method that leverages temporal correlation among sequentially acquired CSI, in conjunction with the proposed VQ structures. Specifically, we propose two types of VQ structures—{\em parallel} VQ and {\em unified} VQ—that not only exploit temporal correlation but also address the issue of compensating for the quantization error from the previous feedback.

    \subsection{Initial Feedback Method: DL-based CSI Feedback with VQ} \label{Sec: VQ-VAE}
\begin{figure}[t]
    \centering 
    
    {\epsfig{file=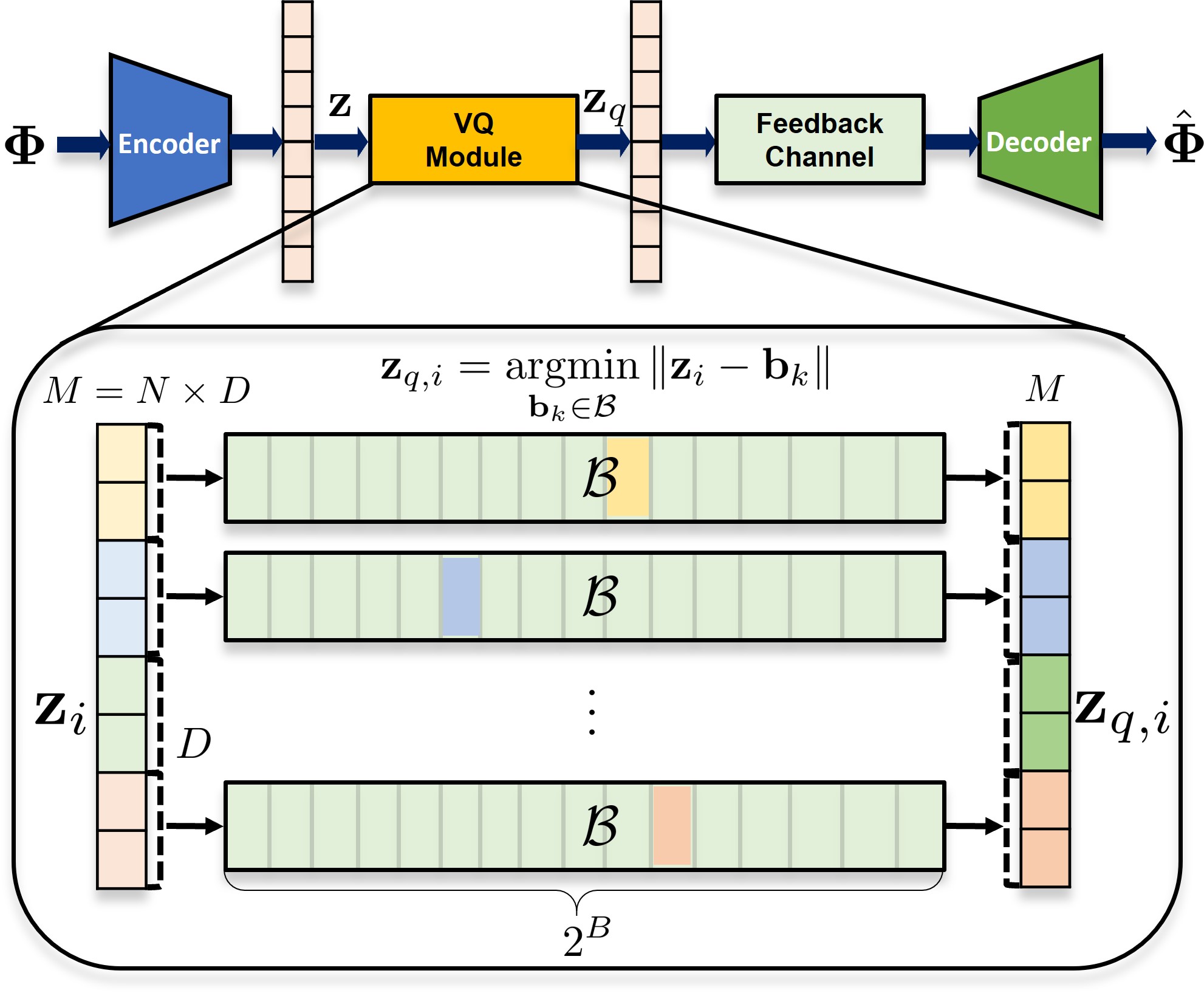, width=7cm}}

    \caption{An illustration of our DL-based initial CSI feedback with VQ.}
    \label{fig: VQVAE}\vspace{-3mm}
    \end{figure}

    Building on the effectiveness of VQ in facilitating finite-bit representations of latent vectors, we adopt it as the foundation of the inital CSI feedback framework.
    Our framework utilizes a trainable quantization codebook placed in the latent space and jointly trains the encoder, codebook, and decoder using a loss function that captures both the quantization and reconstruction errors. 
     To avoid a significant computational complexity required for VQ, we adopt the product VQ structure in \cite{productVQ, ESC-MVQ}. Specifically, we first divide the latent vector ${\bf z}$ into $N$ sub-vectors each with dimension $D$ ($M=N\times D$) and then use a $D$-dimensional codebook for quantizing each sub-vector separately, as illustrated in Fig. \ref{fig: VQVAE}.
    Let $\mathcal{B}$ be a vector codebook using $B$ bits which consists of $2^B$ $D$-dimensional codewords, namely $\{\mathbf{b}_k\}^{2^B}_{k=1}$. 
    Also, let ${\bf z}_i$ be the $i$-th sub-vector of ${\bf z}$, defined as ${\bf z}_i = [z_{(i-1)D+1},\cdots,z_{i D}]$, where $z_j$ is the $j$-th entry of ${\bf z}$.
    Then each sub-vector ${\bf z}_i$ is quantized to $\mathbf{z}_{q,i}$ using the codebook $\mathcal{B}$ according to the minimum Euclidean distance criterion, i.e.,
    \begin{align}
    \mathbf{z}_{q,i}=\argmin_{\mathbf{b}_k \in \mathcal{B}}\Vert\mathbf{z}_i - \mathbf{b}_k \Vert.
    \end{align}
    For jointly training the encoder, codebook, and decoder, we adopt the loss function in \cite{VQ-VAE}, given by 
    \begin{align}\label{eq: VQ-VAE Loss}
    \mathcal{L}_{\text{vq}} = d(\mathbf{X},\mathbf{\hat X})+\left\Vert \text{sg}(\mathbf z)-\mathbf{z}_q\right\Vert^2 + \beta \left\Vert\mathbf z-\text{sg}(\mathbf{z}_q)\right\Vert^2,
    \end{align}
    where $d(\cdot,\cdot)$ is a distortion measure which calculates the difference between two inputs with a certain criterion. Also, sg($\cdot$) is the `stop-gradient' operator, which treats an input as a constant, allowing it to be ignored in gradient descent procedures with a given loss structure. In particular, the last two terms in \eqref{eq: VQ-VAE Loss} depend on a quantization loss, aiming to minimize the difference between $\mathbf{z}$ and $\mathbf{z}_q$.
    The third term in \eqref{eq: VQ-VAE Loss} is called a {\em commitment loss} which is regularized by a hyperparameter $\beta$ \cite{VQ-VAE}. After calculating the quantization loss, a gradient correction of the decoder input is performed as $\mathbf{z}_q\leftarrow\mathbf{z}+\text{sg}(\mathbf{z}_q-\mathbf{z})$.

    In our method, we choose an angle parameter $\mathbf{\Phi}\in \mathbb{R}^{2\times N_a \times N_c}$ as the input $\mathbf{X}$ of the encoder network, which consists of the angle parameters $\{\phi_{i,j}\}$ and $\{\psi_{l,i}\}$, where $N_a={\rm max}( \sum^{N_t-1}_{i={\rm max}(N_t-N_s,1)}i ,1)$ represents the number of each angle parameter per subcarrier.
    For implementation flexibility, we also allow $\mathbf{\Phi}$ to be reshaped into a flattened format $\mathbf{\Phi}\in \mathbb{R}^{1\times 2N_a \times N_c}$, where the two components are concatenated along the angle parameter dimension.
    We aggregate the angle parameters in $\mathbf{\Phi}$ in the same order as specified in the IEEE 802.11 standard.
    The rationale behind this input choice is that we can exploit the abstracted information in the beamforming matrix with orthogonal columns, while effectively accounting for the correlations between these elements. In contrast, using the full matrix $\mathbf{V}$ as the input typically results in a much higher input dimension and requires additional enforcement to ensure orthogonality in the output.
    This rationale is numerically justified in Sec. \ref{Phi input}.
    Ultimately, the proposed DL-based CSI feedback method replaces the uniform quantization used in the conventional standard with an encoder, VQ module, and decoder following the angle parameter extraction process, as illustrated in Fig. \ref{fig: overallflow}.
    
        {\bf Remark (Comparison with Existing DL-based CSI Feedback):}
    Our approach is differentiated from existing DL-based CSI feedback methods for Wi-Fi systems, which are those in \cite{LB-SciFi, EFNet}, with several viewpoints. First, we employ a trainable VQ module which facilitate an efficient finite-bit representation of the latent vector compared to the quantization techniques in  \cite{LB-SciFi, EFNet}. We also consider the angle parameters $\{\phi_{i,j}\}$ and $\{\psi_{l,i}\}$ as an input for the encoder network, resulting in a more efficient input representation than the inputs considered in the existing methods.

    \begin{figure}[t]
    \centering 
    {\epsfig{file=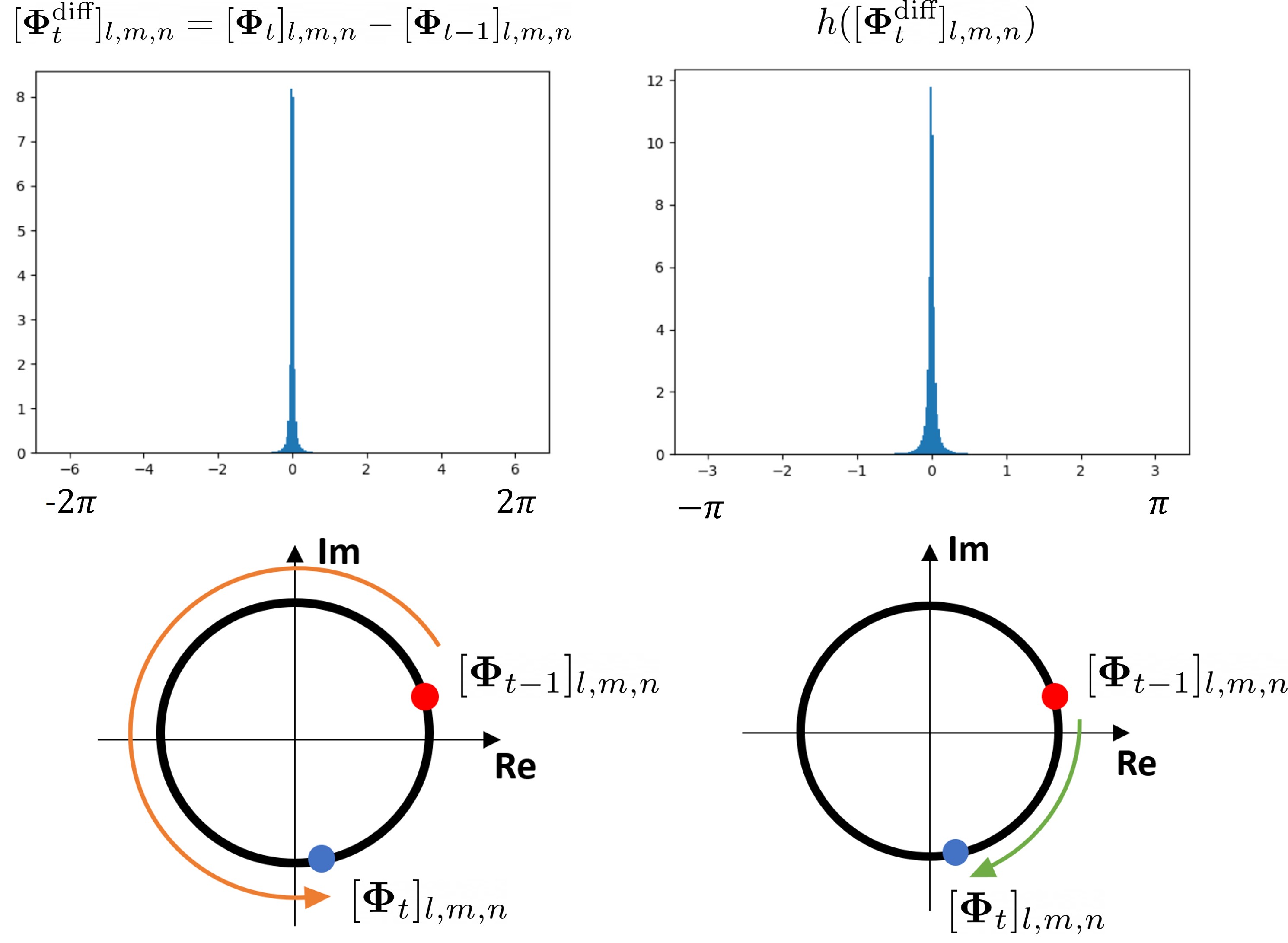, width=7.5cm}}

    \caption{An illustration of preprocessing on $\mathbf{\Phi}^{\rm diff}_{t}$ designed to capture the periodicity of angle parameters.}
    \label{fig: periodicity}\vspace{-3mm}
    \end{figure}

    \subsection{Core Idea: Angle-Difference Feedback}\label{subSec: Time sparisty}
    In practical systems, the STA connected to the AP periodically feeds the CSI back to the AP. In this scenario, the CSIs continuously evolve over time and therefore have temporal correlations up to a certain level. Inspired by this fact, in this subsection, we present a core idea of angle-difference feedback strategy which further enhances the feedback efficiency by exploiting the temporal correlation of the CSIs.
    
    A simple yet effective approach to exploit the temporal correlation of the CSIs is to feed back the difference between the current and previous angle parameters to the AP. If these angle parameters are temporally correlated, their difference would be sparse. This sparsity makes it more efficient to compress the sparse matrix and then sum it with the original matrix after reconstruction. To incorporate this idea into our CSI feedback framework in Sec. \ref{Sec: VQ-VAE}, we consider two feedback phases: (i) an initial CSI feedback phase, and (ii) an angle-difference feedback phase. In the initial feedback phase, the angle parameter $\mathbf{\Phi}_t$ at time $t$ is compressed using an initial encoder, referred to as the Type-I encoder  $f_{\rm enc1}$. The corresponding decoder, referred to as the Type-I decoder $f_{\rm dec1}$, is then utilized at the AP to reconstruct the initial angle parameter. In the angle-difference feedback phase,  the angle difference at time $t$, defined as $\mathbf{\Phi}^{\rm diff}_{t} = \mathbf{\Phi}_{t}-\mathbf{\Phi}_{t-1}$, is compressed using an angle-difference encoder, namely Type-II encoder $f_{\rm enc2}$. The corresponding decoder, referred to as the Type-II decoder $f_{\rm dec2}$, is used to reconstruct the angle difference $\mathbf{\Phi}^{\rm diff}_{t}$. During this phase, the angle parameter is reconstructed by adding the reconstructed angle difference to the previous angle parameter.  

    \subsection{Preprocessing for Angle-Difference Feedback}\label{subSec: preprocessing}
    Using the angle difference $\mathbf{\Phi}^{\rm diff}_{t}$ directly as the input to the Type-II encoder $f_{\rm enc2}$ fails to account for the {\em periodicity} of angle parameters in Wi-Fi systems. To enable the model to effectively capture actual variations in these angle parameters, in this subsection, we design a proper preprocessing to capture the periodicity of the angle parameters in Wi-Fi systems. The goal of the preprocessing function $h(x)$ is to map the angle difference into the minimal rotation range $[-\pi,\pi)$, thereby allowing the model to capture the true angular variation between consecutive time instances more effectively. To achieve this goal, we define a preprocessing function $h(x)$, applied to each element $[\mathbf{\Phi}^{\rm diff}_{t}]_{l,m,n}$, as  
    \begin{align}
    h(x)=\begin{cases}
    x, &\text{if } -\pi\leq x<\pi,\\
    -\text{sgn}(x)\times(2\pi-|x|), &\text{otherwise},
    \end{cases}
    \end{align}
    where $x\in(-2\pi,2\pi)$ and ${\rm sgn}(\cdot)$ represents the sign function, which returns a sign of an input value. The operation of the preprocessing $h(\cdot)$ is illustrated in Fig. \ref{fig: periodicity}, where the angle differences are extracted from the Wi-MIR channel dataset \cite{Wi-MIR}, measured in a real Wi-Fi environment.
    The upper left part of Fig. \ref{fig: periodicity} shows the distribution of $[\mathbf{\Phi}^{\rm diff}_{t}]_{l,m,n}$ without preprocessing. Even though it seems that the majority of elements in $\mathbf{\Phi}^{\rm diff}_{t}$ are concentrated around 0, the values of each element range from $-2\pi$ to $2\pi$. This means that only a counterclockwise rotation is considered to calculate the difference between $\mathbf{\Phi}_{t}$ and $\mathbf{\Phi}_{t-1}$ as illustrated in the lower left part of Fig. \ref{fig: periodicity}. Otherwise, if $\left|[\mathbf{\Phi}^{\rm diff}_{t}]_{l,m,n}\right|$ is greater than $\pi$, a clockwise rotation can be also considered and it yields $-\text{sgn}\big([\mathbf{\Phi}^{\rm diff}_{t}]_{l,m,n}\big)\times\big(2\pi-\left|[\mathbf{\Phi}^{\rm diff}_{t}]_{l,m,n}\right|\big)$ as the angle difference. Eventually, we can determine the angle difference with the smallest rotation, whose absolute value is always smaller than $\pi$, using $h([\mathbf{\Phi}^{\rm diff}_{t}]_{l,m,n})$. The distribution of $h([\mathbf{\Phi}^{\rm diff}_{t}]_{l,m,n})$ is illustrated in the upper right part of Fig. \ref{fig: periodicity}. For $h([\mathbf{\Phi}^{\rm diff}_{t}]_{l,m,n})$, we can see the number of near-zero values of the angle difference has further intensified. In this manner, our preprocessing determines the smallest rotation, which ensures more effective input to consider the temporal correlation of the angle parameters.

    \begin{figure}[t]
        \centering 
        {\epsfig{file=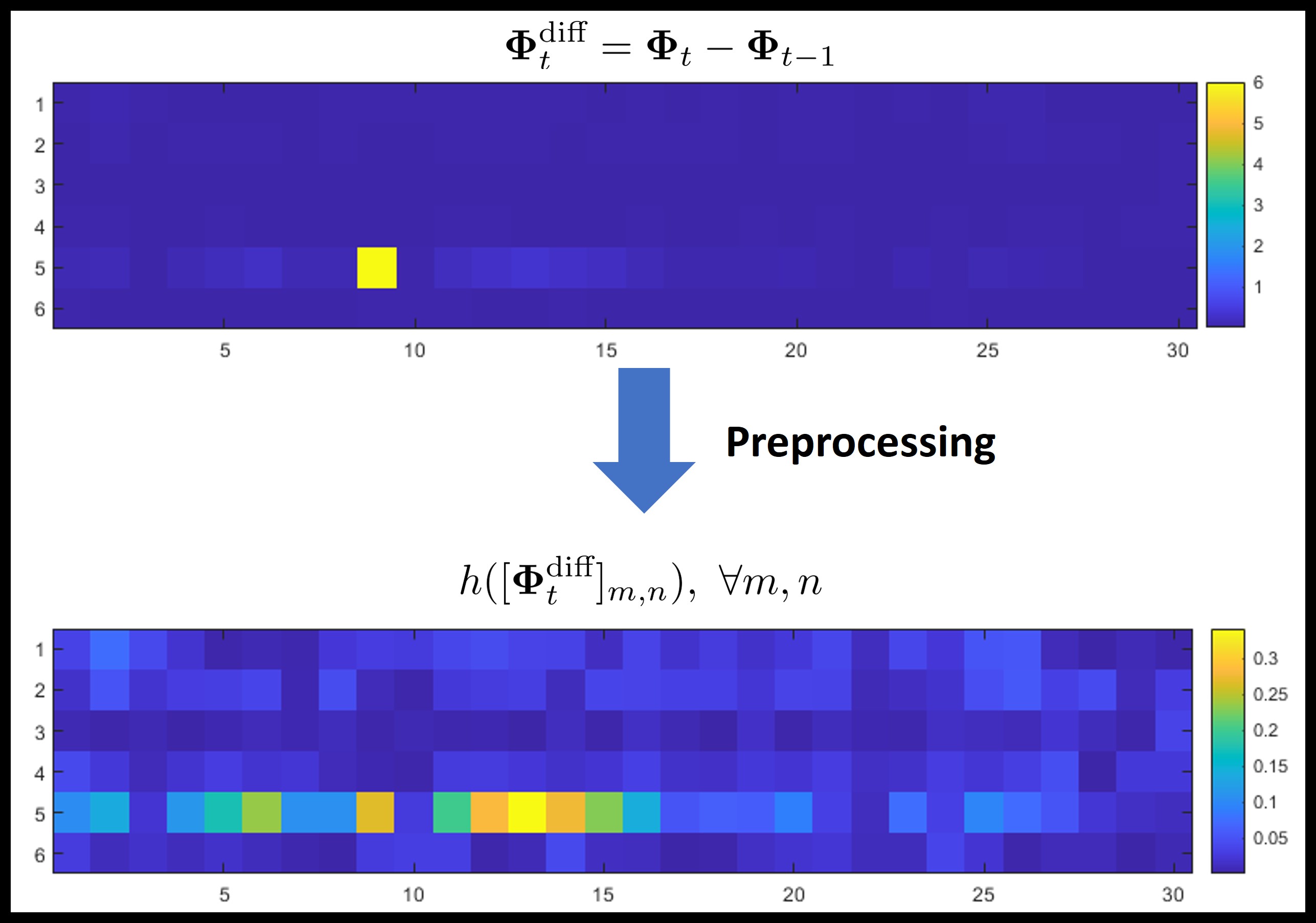, width=7.5cm}}
        \caption{An instance for the scaled image plot of $|[\mathbf{\Phi}^{\rm diff}_{t+1}]_{l,m,n}|$ with and without preprocessing. $(N_r=N_t=N_s=3,N_a=3,N_c=30)$}
    
        \label{fig: scaled_image}\vspace{-3mm}
    \end{figure}

    To illustrate the sparsity of the angle difference with our preprocessing, in Fig.~\ref{fig: scaled_image}, we present an instance for the scaled image plot of the angle difference with and without the our preprocessing. Without the preprocessing, the instance in Fig.~\ref{fig: scaled_image} seems sparse enough, where only a few elements show significant values while the others do not. However, when we apply our preprocessing with $h(\cdot)$, the tendency of the instance clearly differs from that without preprocessing. An instance in Fig. \ref{fig: scaled_image} turns out to be less sparse with the preprocessing. In fact, without the preprocessing,  the most significant element in Fig. \ref{fig: scaled_image} presents the value near $2\pi\ (\approx 6.28)$, which corresponds to near-zero value considering the shortest path. This implies that this value may not be a significant value.

    Building on the preprocessing described above, we design a feedback selection method to determine whether to employ the angle-difference feedback strategy. In this method, the sparsity of the angle difference is evaluated based on the output of the preprocessing step. The angle-difference feedback strategy is selected only if the preprocessed angle difference is sufficiently sparse. To achieve this, we first count the number of non-zero elements at time instance $t$ using a criterion defined as 
    \begin{align}
    N_{d,t} = \sum_{l,m,n}\mathbb{I}\big(\left|h([\mathbf{\Phi}^{\rm diff}_{t}]_{l,m,n})\right| > \mu_{\rm th} \big),
    \end{align}
    where $\mu_{\rm th}$ is a threshold a threshold below which values are considered zero, and $\mathbb{I}(\cdot)$ is the indicator function, returning $1$ if the input condition is true and $0$ otherwise. We then determine that the angle difference is sparse enough if the number of non-zero elements, $N_{d,t}$, is smaller than a threshold $N_{\rm th}$, where $N_{\rm th} < 2N_a N_c$. We denote the decision using an indicator variable  $I_t$ which is set to $1$ if the angle-difference feedback is selected and $0$ otherwise. The indicator variable in our method is determined as
    \begin{align}
    I_{t} = \mathbb{I}(N_{d,t} < N_{\rm th} ).
    \end{align}
    Note that our selection method requires an additional 1-bit feedback to indicate whether the angle-difference feedback is selected or not. Since this additional overhead is only 1 bit, its impact is negligible compared to the overall CSI feedback overhead.
    

    \subsection{Parallel VQ Method for Angle-Difference Feedback}\label{subSec: Multi-Stage VQ}
    During the angle-difference feedback phase, the angle parameter can only be reconstructed by adding the previous angle parameter to the newly received angle difference. Consequently, the reconstruction error from the previous feedback process is inevitably propagated to the current reconstruction process, potentially degrading its performance significantly. To mitigate this error propagation, it is essential to develop an additional error compensation strategy to maximize the overall reconstruction performance during the angle-difference feedback phase.

    To mitigate the error propagation problem, we present a parallel VQ method for our angle-difference feedback strategy. The key idea behind this method is to employ an additional VQ module to quantize the residual error of the CSI feedback. This idea is inspired from a multi-stage VQ structure, which consists of cascade VQ modules compensating the quantization errors sequentially \cite{Multi-stage VQ, Multi-stage VQ_CSI}. In this structure, the subsequent VQ module quantizes the quantization error vector of the previous VQ module. Then, summing the outputs of all VQ modules leads to a better reconstruction performance than using a single-stage VQ. 
    
        \begin{figure}[t]
        \centering 
        {\epsfig{file=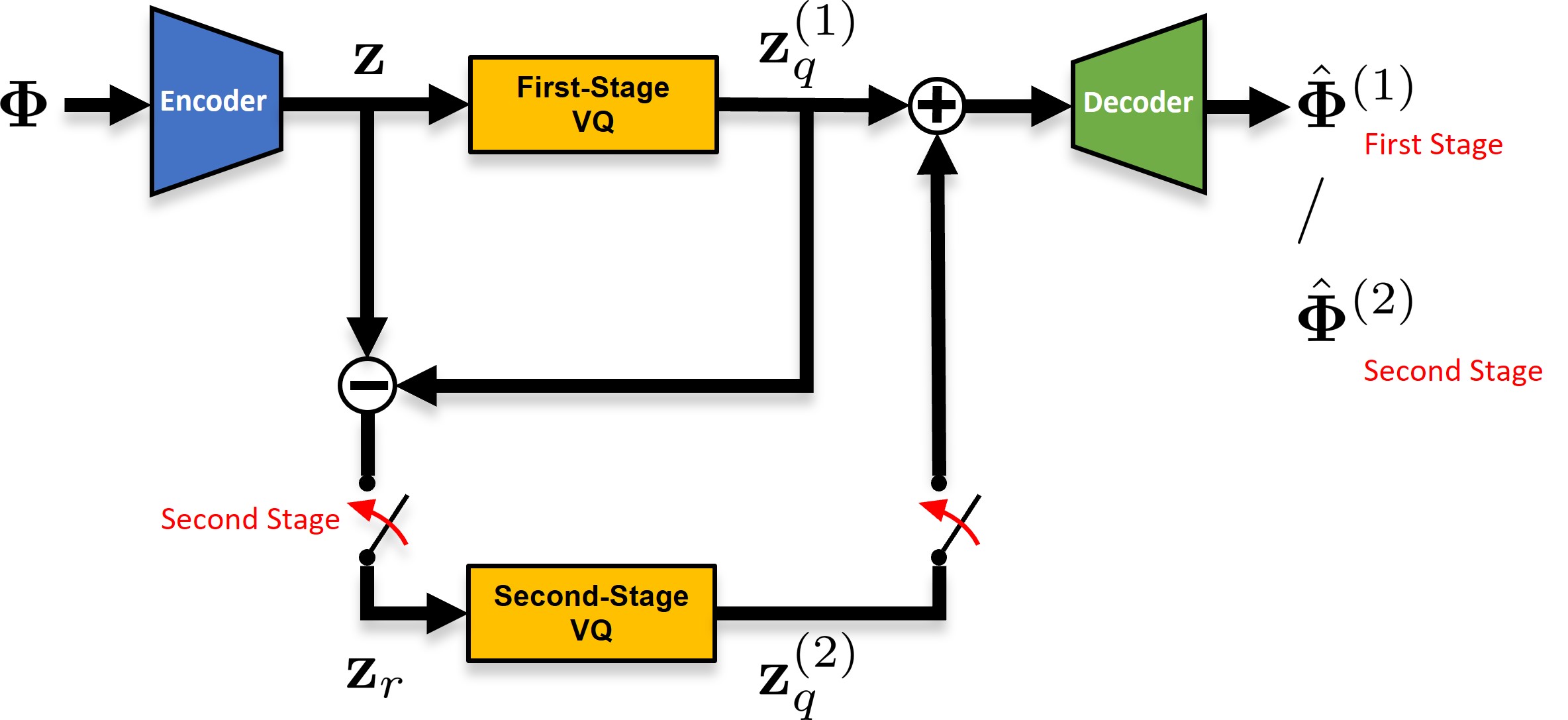, width=7.5cm}}
    
        \caption{A structure of two-stage VQ-based CSI feedback strategy.}
        \label{fig: two-stage}\vspace{-3mm}
    \end{figure}
    
    A two-stage VQ architecture for DL-based CSI feedback is illustrated in Fig. \ref{fig: two-stage}, which is applied to  both  the initial encoder-decoder pair and the angle-difference encoder-decoder pair. Let $\mathbf{z}_q^{(1)}$ be an output of the first-stage VQ module. To enhance the VQ performance, in two-stage VQ structure, the residual latent vector $\mathbf{z}_r=\mathbf{z}-\mathbf{z}_q^{(1)}$ is quantized using the second VQ module. Then, the reconstructed angle parameter for each stage is given by 
    \begin{align}
    \mathbf{\hat \Phi}^{(1)}&=f_{\rm dec}\big(\mathbf{z}_q^{(1)}\big),\\
    \mathbf{\hat \Phi}^{(2)}&=f_{\rm dec}\big(\mathbf{z}_q^{(1)} + \mathbf{z}_q^{(2)}\big),
    \end{align}
    where $\mathbf{\hat \Phi}^{(1)}$ is the angle parameter reconstructed solely from the first-stage VQ module, and $\mathbf{\hat \Phi}^{(2)}$ is the angle parameter reconstructed with both the first-stage and second-stage VQ modules. Since the use of the second-stage VQ module further reduces the quantization error, the reconstruction error of $\mathbf{\hat \Phi}^{(2)}$ is smaller than the reconstruction error of $\mathbf{\hat \Phi}^{(1)}$.


    
    
    
    \begin{figure}[t]
        \centering 
        {\epsfig{file=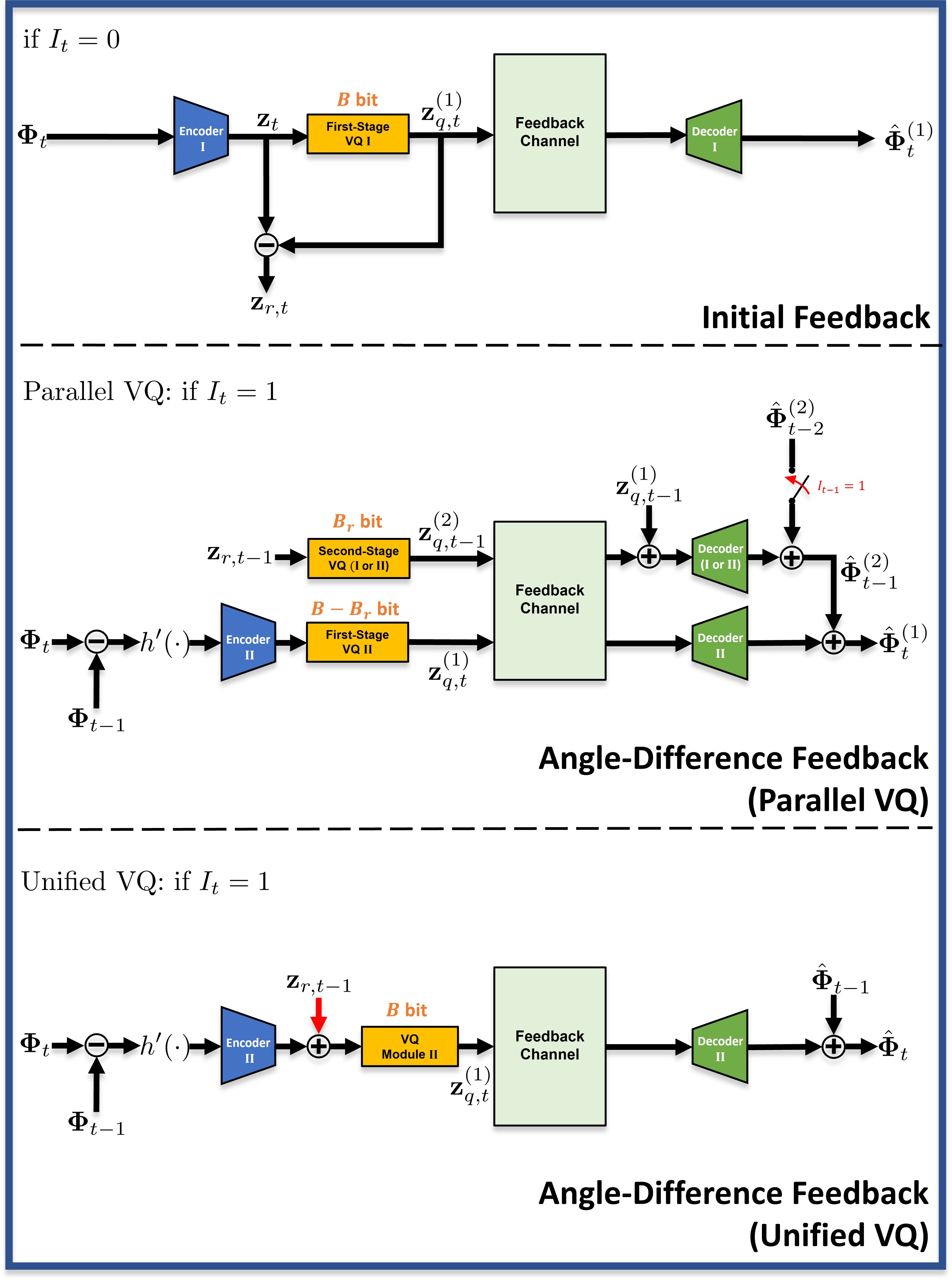, width=9cm}}
    
        \caption{An illustration of VQ frameworks for angle-difference feedback.}
        \label{fig: VQ_strategy_integrated}\vspace{-3mm}
    \end{figure}
    
    In our parallel VQ method, we incorporate the two-stage VQ architecture into the angle-difference feedback strategy by utilizing the second-stage VQ module to mitigate the residual error in the previously-reconstructed angle parameter. To this end, we allocate a portion of the available feedback bits, given by $B_r <B$, to the second-stage VQ module to feed back the quantization of the residual error in first-stage VQ module in the previous feedback process. Meanwhile, we allocation the remaining $B-B_r$ bits to the first-stage VQ module to feed back the information of the current angle difference. Note that during the initial feedback phase, only the first-stage VQ module is utilized because there is no residual error propagated from the previous feedback process. Therefore, in this case, all the $B$ bits are allocated solely to the first-stage VQ module. Our parallel VQ method is illustrated in Fig. \ref{fig: VQ_strategy_integrated}.

    We provide additional details of the feedback process of the STA in our parallel VQ method, illustrated in Fig. \ref{fig: VQ_strategy_integrated}. 
    At time instance $t$, when $I_t=0$ at time instance $t$, the angle parameter $\mathbf{\Phi}_{t}$ is fed back to the AP using the first-stage VQ module. When $I_t=1$, the preprocessed angle difference $h'(\mathbf{\Phi}^{\rm diff}_{t})$ are fed back to the AP, where $h'(\cdot)$ is a function which produces a matrix as an output, satisfying $[h'(\mathbf A)]_{l,m,n}=h([\mathbf A]_{l,m,n}),\forall l,m,n$ with an input matrix $\mathbf{A}$. Simultaneously, the residual latent vector of the previous feedback, denoted as $\mathbf{z}_{r,t-1}=\mathbf{z}_{t-1}-\mathbf{z}^{(1)}_{q,t-1}$, are also fed back to the AP. Specifically, $h'(\mathbf{\Phi}^{\rm diff}_{t})$ is used as input to the Type-II encoder, and its output is quantized by the first-stage VQ module. The quantized output, $\mathbf{z}^{(1)}_{q,t}$, is then fed back to the AP. Concurrently, the residual latent vector $\mathbf{z}_{r,t-1}$ is quantized using the second-stage VQ module, and its quantized output,  $\mathbf{z}^{(2)}_{q,t-1}$,  is also sent to the AP. It is important to note that $B-B_r$ bits are allocated to transmit $\mathbf{z}^{(1)}_{q,t}$, while $B_r$ bits are allocated to transmit $\mathbf{z}^{(2)}_{q,t-1}$. Consequently, the total feedback overhead remains $B$ bits.

    We now describe the details of the reconstruction process at the AP in our parallel VQ method, illustrated in Fig. \ref{fig: VQ_strategy_integrated}. At time instance $t$, when $I_t=0$, the angle parameter $\mathbf{\Phi}_{t}$ is directly reconstructed from the received latent vector $\mathbf{z}^{(1)}_{q,t}$ using the Type-I decoder. When $I_t=1$, the quantized residual latent vector  $\mathbf{z}^{(2)}_{q,t-1}$ received by the AP is added to the previously-received latent vector $\mathbf{z}^{(1)}_{q,t-1}$. From this aggregated vector, $\mathbf{z}^{(1)}_{q,t-1} + \mathbf{z}^{(2)}_{q,t-1}$, a refined angle parameter $\mathbf{\hat \Phi}_{t-1}^{(2)}$ for the previous time instance $t-1$ is reconstructed. In particular, the reconstruction process of $\mathbf{\hat \Phi}_{t-1}^{(2)}$ depends on the type of the previous feedback process. 
    \begin{itemize}
        \item If $I_{t-1}=0$, the previous angle parameter $\mathbf{\Phi}_{t-1}$ is reconstructed from the aggregated latent vector using the Type-I decoder.
        \item If $I_{t-1}=1$, the previous angle difference $\mathbf{\Phi}_{t-1} - \mathbf{\Phi}_{t-2}$ is reconstructed from the aggregated latent vector using the Type-II decoder. Then, the refined angle difference is added to the previously refined angle parameter $\mathbf{\hat \Phi}_{t-2}^{(2)}$ at time instance $t-2$. 
    \end{itemize} 
    Based on these considerations, the refined angle parameter $\mathbf{\hat \Phi}_{t-1}^{(2)}$ is expressed as
    \begin{align}\label{eq:Refined_angle}
        \mathbf{\hat \Phi}_{t-1}^{(2)}=
        \begin{cases}
            f_{\rm dec1}\big(\mathbf{z}^{(1)}_{q,t-1} + \mathbf{z}^{(2)}_{q,t-1}\big),  & \text{if}~I_{t-1}=0,\\
            f_{\rm dec2}\big(\mathbf{z}^{(1)}_{q,t-1} + \mathbf{z}^{(2)}_{q,t-1}\big)+\mathbf{\hat \Phi}_{t-2}^{(2)}, &\text{else if}~I_{t-1}=1.
        \end{cases}
    \end{align}
    After this, the angle difference at time instance $t$ is reconstructed from the latent vector $\mathbf{z}^{(1)}_{q,t}$ using the Type-II decoder. The reconstructed angle difference is added with the refined angle parameter $\mathbf{\hat \Phi}_{t-1}^{(2)}$ in \eqref{eq:Refined_angle}. The final output of the reconstruction process is expressed as
    \begin{align}
    \mathbf{\hat \Phi}^{(1)}_t=
    \begin{cases}
    f_{\rm dec1}\big(\mathbf{z}_{q,t}^{(1)} \big), &\text{if}~I_{t}=0,\\
    f_{\rm dec2}\big(\mathbf{z}^{(1)}_{q,t}\big) + \mathbf{\hat \Phi}_{t-1}^{(2)}, &\text{else if}~I_{t}=1.\\
    \end{cases}
    \end{align}


    To jointly train the two encoder-decoder pairs in our parallel VQ method, we begin by adopting the training method described in Sec. \ref{Sec: VQ-VAE} to initialize the Type-I encoder-decoder pair, using {\em only} the first-stage VQ module. Once this initialization is complete, we connect the second-stage VQ module to the first-stage VQ module of the pre-trained Type-I encoder-decoder pair. This initialization process is crucial for maximizing the performance of the initial feedback phase, particularly when there is no temporal correlation among the CSIs. In contrast to the Type-I encoder-decoder pair, the Type-II encoder-decoder pair is randomly initialized along with its first-stage and second-stage VQ modules, as it is challenging to train this pair without the pre-trained Type-I encoder-decoder pair. After linking the Type-I and Type-II encoder-decoder pairs, each incorporating two VQ modules, as illustrated in Fig. \ref{fig: VQ_strategy_integrated}, we jointly train all pairs and VQ modules using the following loss function:
    \begin{align}\label{eq: Multi-stage VQ-VAE Loss}
        &\mathcal{L}_{\text{parallel}} = \left\Vert {\rm Exp}\big({j\mathbf{\Phi}_t}\big) - {\rm Exp}\big({j\mathbf{\hat{\Phi}}^{(1)}_t}\big)\right\Vert^2_{\rm F}\nonumber\\
        &~~~+\left\Vert \text{sg}(\mathbf{z}_t)-\mathbf{z}^{(1)}_{q,t}\right\Vert^2 + \beta_1 \left\Vert\mathbf{z}_t-\text{sg}\big(\mathbf{z}^{(1)}_{q,t}\big)\right\Vert^2\nonumber\\
        &~~~+\left\Vert \text{sg}(\mathbf{z}_{r,t-1})-\mathbf{z}^{(2)}_{q,t-1}\right\Vert^2 + \beta_2 \left\Vert\mathbf{z}_{r,t-1}-\text{sg}\big(\mathbf{z}^{(2)}_{q,t-1}\big)\right\Vert^2,
    \end{align}
    where ${\rm Exp}(\cdot)$  is a function that outputs a matrix consisting of the exponentials of the corresponding elements in the input matrix.
    Specifically, the distortion measure in the loss is chosen to be $d(\mathbf{\Phi},\mathbf{\hat \Phi})=\big\Vert {\rm Exp}({j\mathbf{\Phi}}) - {\rm Exp}({j\mathbf{\hat{\Phi}}})\big\Vert^2_{\rm F}$, in order to capture the periodicity and temporal correlation of the angle parameters in Wi-Fi systems. This distortion measure ensures the shortest path in minimizing the difference between the model output and true value.


    \subsection{Unified VQ Method for DL-based Angle-Difference Feedback}\label{subSec: Time sparisty VQ}
    

    
    The parallel VQ method described in the previous subsection effectively leverages the temporal correlation of the CSIs but presents challenges in terms of implementation and optimization. On the one hand, this method requires the Type-I decoder and Type-II decoder to operate in parallel during the angle-difference feedback phase, which increases the computational burden at the AP. On the other hand, the performance of the parallel VQ method depends heavily on the hyperparameter $B_r$, which determines the number of feedback bits allocated to compensate for the previous feedback error. Optimizing this parameter is challenging because it is highly dependent on the channel distribution and the degree of temporal correlation. Intuitively, using a fixed value of $B_r$ across all scenarios is likely to be suboptimal.


    To effectively exploit the temporal correlation of the CSIs while avoiding the aforementioned challenges, we propose a method referred to as the unified VQ method for angle-difference feedback, as illustrated in Fig. \ref{fig: VQ_strategy_integrated}. To provide additional details of the feedback process, suppose that $I_t=1$ at time instance $t$. In the unified VQ method, no additional VQ modules are incorporated into the Type-I or Type-II encoder-decoder pairs. Instead, to compensate for the quantization error from the previous feedback, the residual latent vector $\mathbf{z}_{r,t-1}$ is directly added to the current encoder output $f_{\rm enc2}(h'(\mathbf{\Phi}^{\rm diff}_{t}))$. The sum of $\mathbf{z}_{r,t-1}$ and $f_{\rm enc2}(h'(\mathbf{\Phi}^{\rm diff}_{t}))$ is then quantized by a single unified VQ module using $B$ bits. As a result, in this method, the effective latent vector  $\mathbf{z}_t$, which serves as the input to the VQ module, is expressed as
    \begin{align}\label{eq: residualz_t}
    \mathbf{z}_t=
    \begin{cases}
        f_{\rm enc1}(\mathbf{\Phi}_t),  &\text{if}~I_t=0,\\
        f_{\rm enc2}(h'(\mathbf{\Phi}^{\rm diff}_{t})) + \mathbf{z}_{r,t-1}, &\text{else if}~I_t=1,
    \end{cases}
    \end{align}
    As described above, our unified method not only simplifies the feedback process but also eliminates the need for the hyperparameter $B_r$, as the unified VQ module implicitly accounts for the effects of both the residual latent vector and the current latent vector. This design enables us to avoid optimizing $B_r$ based on the channel distribution or the degree of temporal correlation.

    The reconstruction process at the AP in our unified method is also straightforward. Upon the reception of the feedback, the angle parameter or the angle difference is directly reconstructed from the quantized latent vector $ \mathbf{z}_{q,t}$ as follows:
    \begin{align}\label{eq: sparse reconstruction}
    \mathbf{\hat \Phi}_t =
    \begin{cases}
    f_{\rm dec1}(\mathbf{z}_{q,t}),~~~~~~~~~~~~\text{if }I_t=0,\\
    f_{\rm dec2}(\mathbf{z}_{q,t}) + \mathbf{\hat \Phi}_{t-1},~~\text{else if }I_t=1.\\
    \end{cases}
    \end{align}
    The Type-II decoder in this method can be viewed as a unified version of the parallel decoders used in our parallel VQ method. By leveraging this single unified decoder, the need to run two separate decoders is eliminated, thereby reducing the computational complexity of the reconstruction process during the angle-difference feedback phase, compared to the parallel VQ method in Sec. \ref{subSec: Multi-Stage VQ}. 
    


    To train the two encoder-decoder pairs in our unified VQ method, we adopt the training method described in Sec. \ref{Sec: VQ-VAE} to initialize the Type-I encoder-decoder pair, while randomly initializing the Type-II encoder-decoder pair. Once this initialization is complete, we link the Type-I and Type-II encoder-decoder pairs, as illustrated in Fig. \ref{fig: VQ_strategy_integrated}, and then jointly train these pairs using the following loss function:
    \begin{align}\label{eq: Sparse VQ-VAE Loss}
    \mathcal{L}_{\text{unified}} =& \left\Vert {\rm Exp}({j\mathbf{\Phi}_t}) - {\rm Exp}({j\mathbf{\hat{\Phi}}_t})\right\Vert^2_{\rm F}\nonumber\\&+\left\Vert \text{sg}(\mathbf{z}_t)-\mathbf{z}_{q,t}\right\Vert^2 + \beta \left\Vert\mathbf{z}_t-\text{sg}(\mathbf{z}_{q,t})\right\Vert^2.
    \end{align}

    \section{Proposed DL-based CSI Refinement Framework}\label{Sec:Refine}
    In Sec. \ref{Sec: AD}, we have focused on an STA's strategy for exploiting the temporal correlation of CSIs. These correlations, in fact, can also be utilized on the AP side to further enhance the CSI reconstruction performance. Motivated by this, in this section, we propose a novel DL-based CSI refinement framework  for Wi-Fi systems with temporally-correlated channels, designed to refine the reconstructed angle parameters by jointly leveraging historical angle parameters. 

    \begin{figure}[t]
        \centering 
        {\epsfig{file=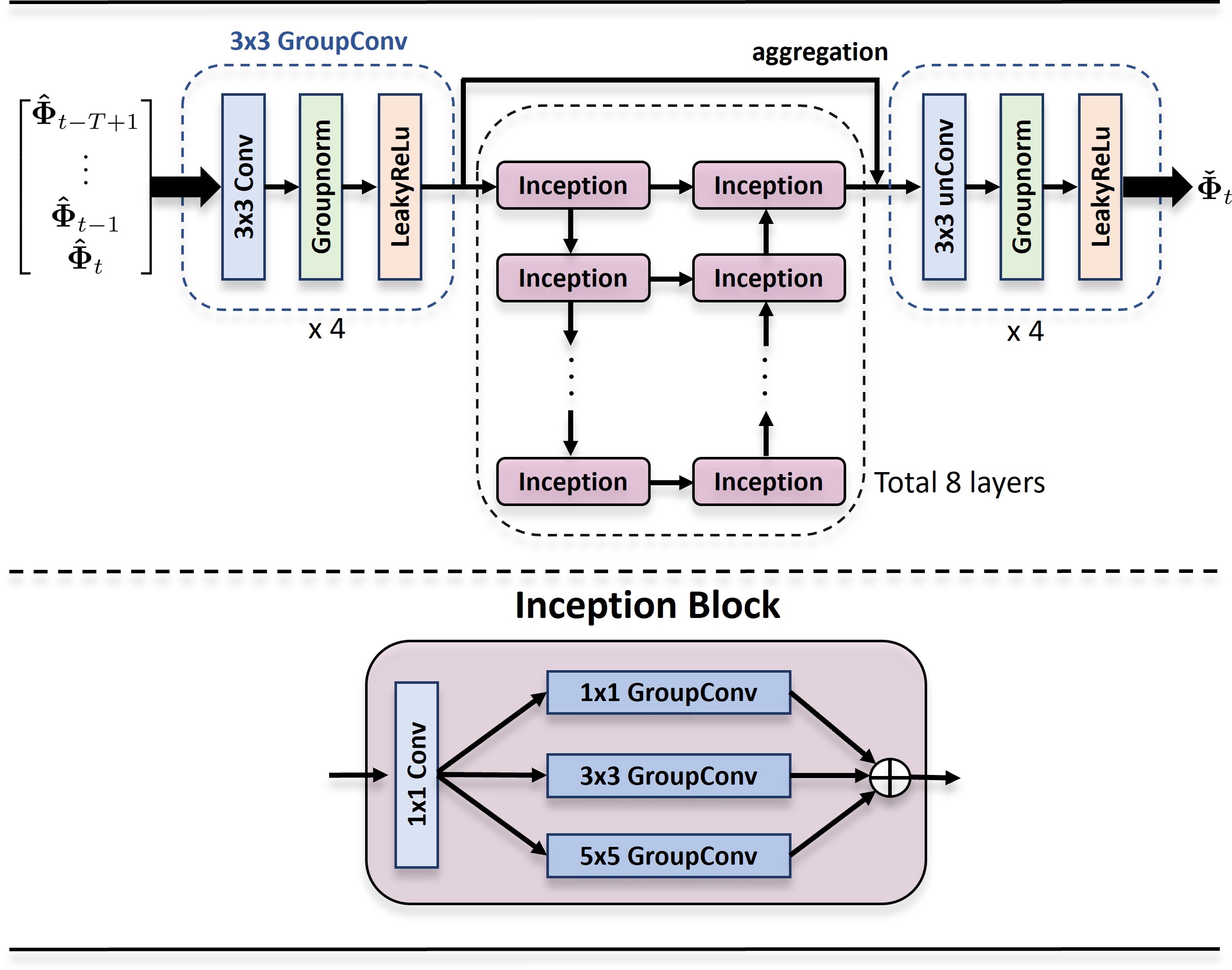, width=7cm}}
        \caption{An illustration of overall structure of proposed CSI refinement module.}
        \label{fig: SimVP}\vspace{-3mm}
    \end{figure}
    
    \subsection{CSI Refinement Module} \label{subSec:Refine_simple}
    In our framework, we introduce a CSI refinement module  which exploits received CSI feedback to reconstruct the current CSI more accurately. This module is followed by the CSI reconstruction process and therefore applied at the AP side only. 
    Our CSI refinement module utilizes the most recently reconstructed $T$ angle parameters $\mathbf{\hat \Phi}_{t-T+1:t}$ to refine the current angle parameter $\mathbf{\hat \Phi}_t$ as
    \begin{align}\label{eq: refinement}
    f_{\rm ref}\big(\mathbf{\hat \Phi}_{t-T+1},...,\mathbf{\hat \Phi}_{t-1}, \mathbf{\hat \Phi}_{t}\big) = \mathbf{\check \Phi}_{t},
    \end{align}
    where $f_{\rm ref}$ represents the CSI refinement module and $\mathbf{\check \Phi}_{t}$ represents the refined angle parameter at time instance $t$. 
    Note that the input to our CSI refinement module also includes the {\em current} angle parameter reconstructed at time instance $t$, which is why our module is considered a {\em refinement} module. Nevertheless, our module can be redesigned as a {\em prediction} module by excluding the current angle parameter from its input.
    For a neural network design of our CSI refinement module, $f_{\rm ref}(\cdot)$, we adopt the video prediction model proposed in \cite{SimVP}, which was originally developed to predict future frames based on recent frames in a video. The primary reason for this choice is that the model features a simple CNN-based architecture with relatively few computational complexities, while delivering fair performance, as will be demonstrated in Sec. \ref{subSec: refine_performance}. The neural network design for our refinement module is illustrated in Fig.~\ref{fig: SimVP}.


    Now, we explain a training strategy for our CSI refinement module. It is important to note that if the CSI feedback and refinement modules are trained jointly, the performance of the CSI feedback module may not be guaranteed when the CSI refinement module is not activated. To avoid this problem, we train our CSI refinement module after completing the training of the aforementioned CSI feedback model. Specifically, we freeze the parameters of the encoder, decoder, and VQ modules after their training and use the outputs of the decoder during the training of the CSI refinement module. The loss function for training the CSI refinement module is defined as
    \begin{align}\label{eq: refinement loss}
        \mathcal{L}_{\rm refine} = \left\Vert {\rm Exp}({j\mathbf{\Phi}}_t) - {\rm Exp}\left({j\mathbf{\check{\Phi}}_t}\right)\right\Vert^2_{\rm F}.
    \end{align}
    As explained in Sec.~\ref{subSec: Multi-Stage VQ}, this loss function is designed to capture both the temporal correlation and the periodicity of the angle parameters. Our training approach allows for the opportunistic activation of the refinement module, enabling the AP to decide whether or not to use the module depending on available memory or computing resources.

    \begin{figure}[t]
        \centering 
        {\epsfig{file=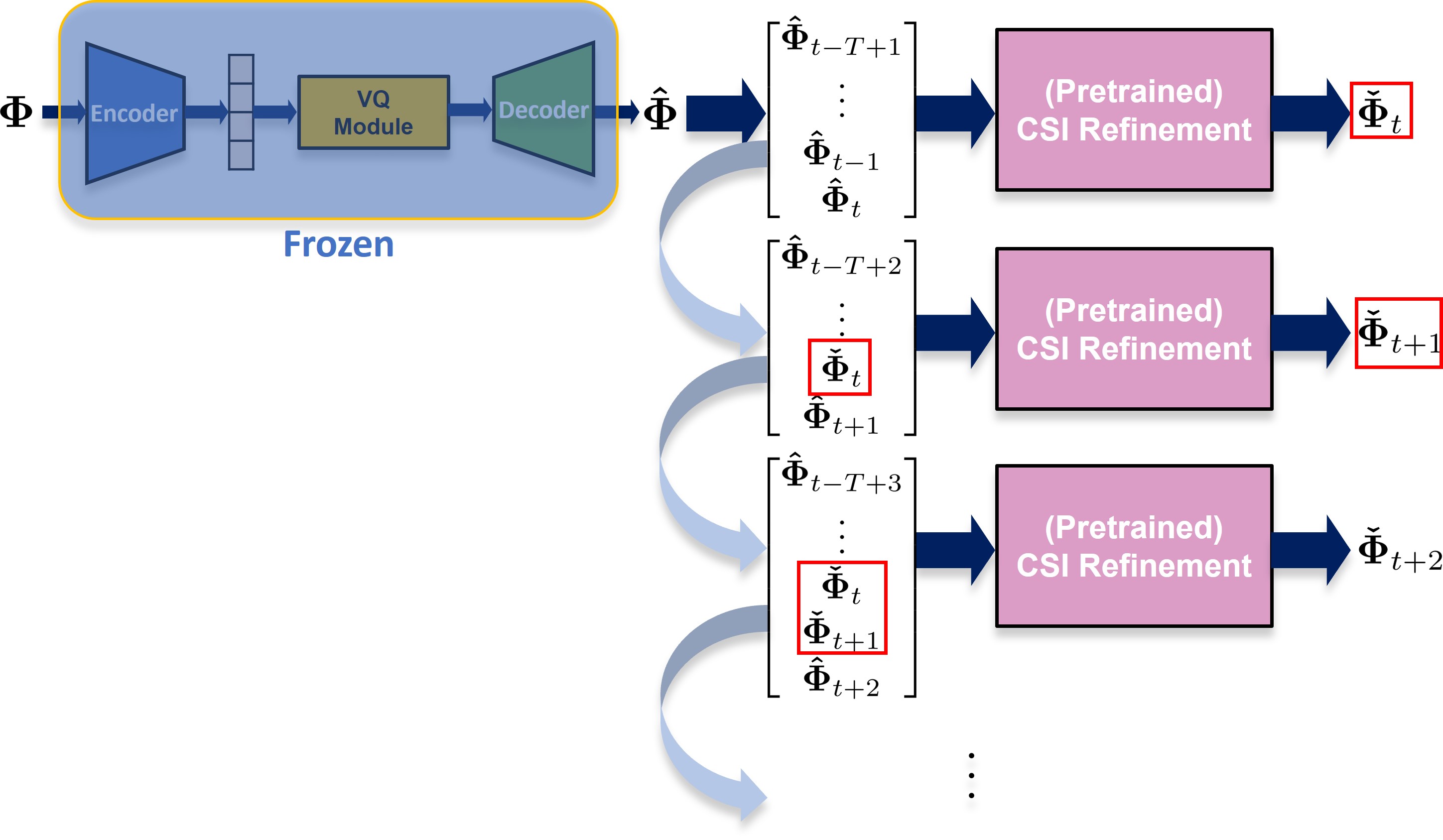, width=7.5cm}}
        \caption{An illustration of the proposed recursive refinement strategy.}
        \label{fig: Refinement_recur}\vspace{-3mm}
    \end{figure}

    \subsection{Recursive Refinement Strategy}
    To enhance the performance of the CSI refinement process, we devise a recursive refinement strategy, in which the refined angle parameter $\mathbf{\check \Phi}_t$, instead of the originally reconstructed parameter $\mathbf{\hat \Phi}_t$, is utilized as an input for the subsequent CSI refinement processes. Our recursive strategy is illustrated in Fig. \ref{fig: Refinement_recur}. 
     In this strategy, we initially train the CSI refinement module over the initial $E_{\rm pre}$ epochs according to the relationship in \eqref{eq: refinement} and the loss function in \eqref{eq: refinement loss}. Following these initial epochs, we begin a recursive training approach with time index $t>T$ by setting
    \begin{align}\label{eq: refinement_recur}
    \mathbf{\check \Phi}_{t}
    \!=\! 
    \begin{cases}
    f_{\rm ref}\big(\mathbf{\check \Phi}_{t-T+1},...,\mathbf{\check \Phi}_{t-1}, \mathbf{\hat \Phi}_{t}\big), &\!\!\!\!\!\!\!\!\!\!\!\!\! \text{if}~t\geq 2T-1,\\
    f_{\rm ref}\big(\mathbf{\hat \Phi}_{t-T+1},...,\mathbf{\hat \Phi}_{T-1},\mathbf{\check \Phi}_{T},...,\mathbf{\check \Phi}_{t-1}, \mathbf{\hat \Phi}_{t}\big), &\!\!\!   \text{otherwise},
    \end{cases}
    \end{align}
    with gradients of all refined CSI inputs corrected by $\mathbf{\check \Phi}\leftarrow {\rm sg}(\mathbf{\check \Phi})$. 
    Since the refined angle parameter is expected to be more accurate than the reconstructed one, our recursive strategy can further enhance the accuracy of the refined angle parameter.

    \section{Simulation Results}\label{Sec:Simul}
    In this section, using simulations, we demonstrate the superiority of the proposed CSI feedback and refinement frameworks for Wi-Fi systems with temporally correlated channels.

    \begin{figure}[t]
        \centering 
        
        {\epsfig{file=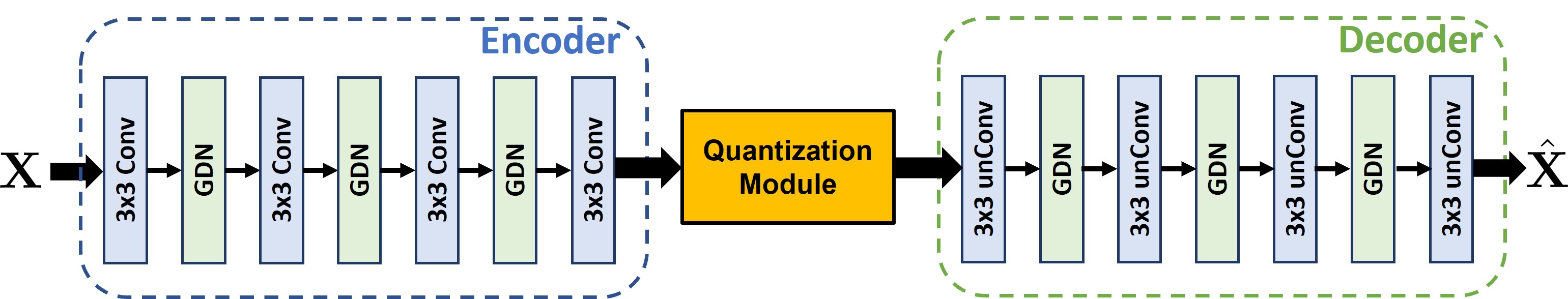, width=8cm}}
    
        \caption{An illustration of encoder and decoder neural network structures, where GDN stands for generalized divisive normalization.}
        \label{fig: CNN2}
    \end{figure}
        
    
    \begin{table}[t]
    \centering
    \caption{Simulation parameter settings for the DeepMIMO channel dataset.}\label{table: configuration}
    \begin{tabular}{|c||c|}
    \hline
    \textbf{Scenario} & DeepMIMO I3 Scenario \\ \hline
    \textbf{Antenna configuration $(N_t,N_r,N_s)$} & $(8,2,2)$ \\ \hline
    \textbf{Number of angle parameter $(N_a)$} & $13$ \\ \hline
    \textbf{Operating frequency band} & 2.4 GHz \\ \hline
    \textbf{System bandwidth} & 20 MHz \\ \hline
    \textbf{Maximum number of channel paths} & 25 \\ \hline
    \textbf{Total number of subcarriers} & 256 \\ \hline
    \textbf{Number of sampled subcarriers $(N_c)$} & 64 \\ \hline
    \textbf{Size of (train/val./test) dataset} & (80,000/15,000/15,000) \\ \hline
    \textbf{Batch size} & 200 \\ \hline
    \textbf{Feedback frequency} & 40 Hz \\ \hline
    \textbf{Speed of STAs} & 0.4 m/s or 0.56 m/s  \\ \hline
    \end{tabular}
    \end{table}

    \begin{table}[t]
    \centering
    \caption{Simulation parameter settings for the Wi-MIR channel dataset.}\label{table: configuration2}
    \begin{tabular}{|c||c|}
    \hline
    \textbf{Antenna configuration $(N_t,N_r,N_s)$} & $(3,3,3)$ \\ \hline
    \textbf{Number of angle parameter $(N_a)$} & $3$ \\ \hline
    \textbf{Operating frequency band} & 5 GHz \\ \hline
    \textbf{System bandwidth} & 20 MHz \\ \hline
    \textbf{Total number of subcarriers} & 64 \\ \hline
    \textbf{Number of sampled subcarriers $(N_c)$} & 30 \\ \hline
    \textbf{Size of (train/val./test) dataset} & (100,000/30,000/30,000) \\ \hline
    \textbf{Batch size} & 200 \\ \hline
    \textbf{Feedback frequency} & 23.75 Hz \\ \hline
    \end{tabular}
    \end{table}

    \subsection{Simulation Setting}
    In our simulations, we employ two channel datasets for training and testing: the DeepMIMO channel dataset on the I3 scenario \cite{DeepMIMO}, and the Wi-MIR channel datset \cite{Wi-MIR}. The DeepMIMO dataset is synthetically generated following the IEEE 802.11be standards \cite{Wi-Fi7}. The Wi-MIR dataset is measured in a real Wi-Fi environment using the IEEE 802.11 CSI tool platform in \cite{CSI tool}, which has also been adopted in \cite{LB-SciFi,EFNet}. The key parameter settings for each dataset are summarized in Table~\ref{table: configuration} and Table~\ref{table: configuration2}, respectively.
    The parameters related to the proposed frameworks are set as $D=16$, $B=8$, and $\beta_1=\beta_2=\beta=0.25$. Also, the threshold parameters are set to $(\mu_{\rm th},N_{\rm th})=(0.3\pi,20)$ for the DeepMIMO dataset and $(\mu_{\rm th},N_{\rm th})=(0.1\pi,5)$ for the Wi-MIR dataset.
    The proposed DL-based CSI feedback and refinement models are implemented and trained using PyTorch 1.12.0 in a Python 3.9.8 environment. The CSI feedback model is trained using the Adam optimizer with $10^{-4}$ learning rate for 1000 epochs. The CSI refinement module is pretrained with $10^{-3}$ learning rate during the initial $E_{\rm pre}=80$ epochs. Then, this module is trained using a recursive training approach with $10^{-4}$ learning rate for the subsequent 2000 epochs.
    The neural network structures for encoder and decoder are constructed as in Fig. \ref{fig: CNN2}. Note that the same structures are utilized for both the Type-I and Type-II encoder and decoders for the proposed angle-difference feedback framework in Sec. \ref{subSec: Multi-Stage VQ} and Sec. \ref{subSec: Time sparisty VQ}.
    Although we have specified the simple neural network architecture for implementing the DL-based CSI feedback model, the proposed methods do not commit to a specific encoder–decoder structure for CSI feedback. 
    For the performance measure, we adopt the normalized mean squared error (NMSE) defined as
    $\mathbb{E}\big\{{\Vert\overline{\mathbf V}-\hat{\mathbf V}\Vert^2_{\text F}}/{\Vert\overline{\mathbf V}\Vert^2_{\text F}}\big\}$ or $\mathbb{E}\big\{{\Vert{\mathbf V}-{\mathbf V}_{\mathbf{\hat \Phi}}\Vert^2_{\text F}}/{\Vert{\mathbf V}\Vert^2_{\text F}}\big\}$ for each input where ${\mathbf V}_{\mathbf{\hat \Phi}}$ represents $\mathbf{V}$ derived from reconstructed angles using \eqref{eq: angle_to_V}\footnote{Note that two kinds of definitions are exactly equivalent, given that the equations $\big\Vert{\mathbf V}\big\Vert^2_{\text F}=\big\Vert\overline{\mathbf V}\tilde{\mathbf D}^*\big\Vert^2_{\text F}=\big\Vert\overline{\mathbf V}\big\Vert^2_{\text F}$ and $\big\Vert{\mathbf V}-{\mathbf V}_{\mathbf{\hat \Phi}}\big\Vert^2_{\text F}=\big\Vert{\mathbf V}\tilde{\mathbf D}^*-{\mathbf V}_{\mathbf{\hat \Phi}}\tilde{\mathbf D}^*\big\Vert^2_{\text F} = \big\Vert\overline{\mathbf V}-{\mathbf V}_{\mathbf{\hat \Phi}}\tilde{\mathbf D}^*\big\Vert^2_{\text F}$ hold. Then ${\mathbf V}_{\mathbf{\hat \Phi}}\tilde{\mathbf D}^*$ corresponds to $\mathbf{\hat V}$.}.

    \begin{figure}[t]
        \centering 
        
        \subfigure[DeepMIMO]{\epsfig{file=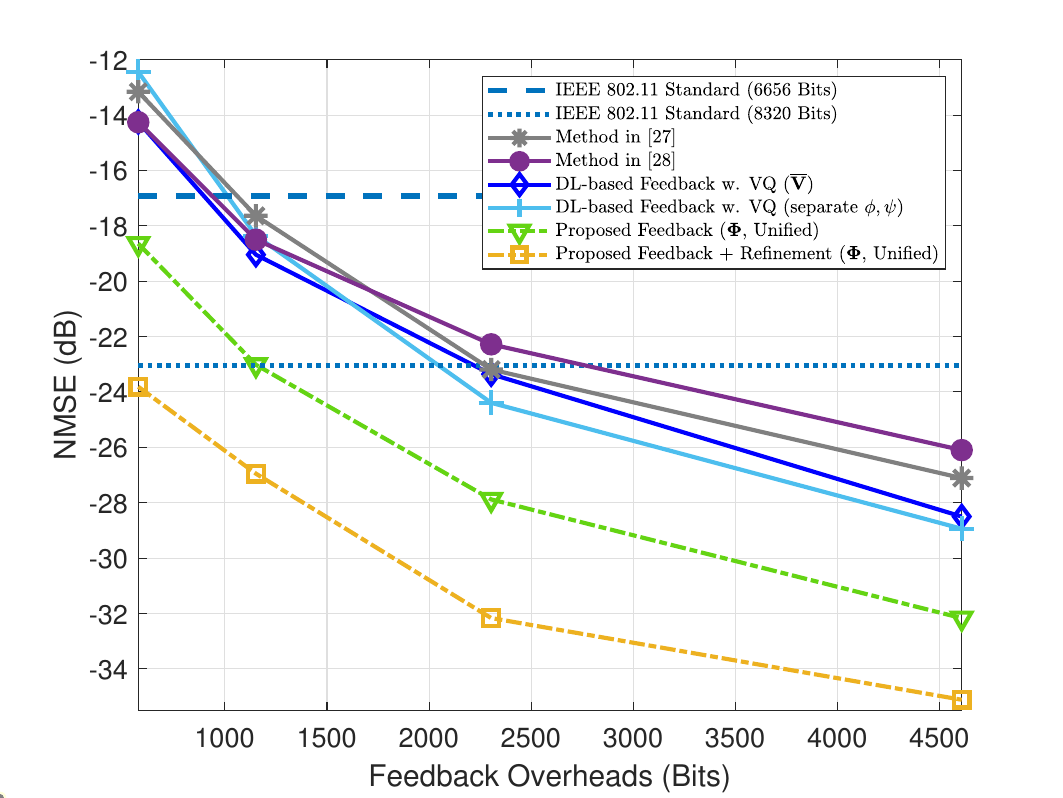, width=7.5cm}}

        \subfigure[Wi-MIR]
        {\epsfig{file=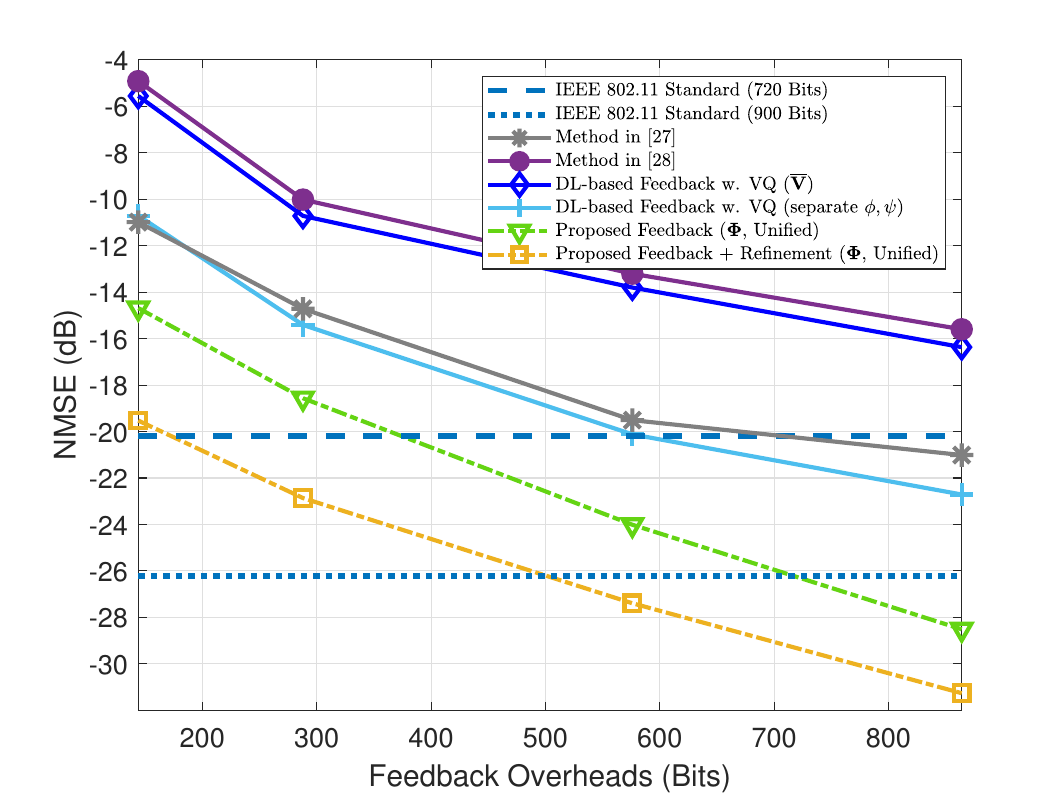, width=7.5cm}}
    
        \caption{Comparison of the NMSE (dB) performance of the CSI feedback framework with various input types and quantization techniques.}
        \label{fig: V_vs_angle}\vspace{-3mm}
    \end{figure}

    \subsection{NMSE Performance of the Proposed Frameworks}\label{Phi input}
    In this subsection, we demonstrate the superiority of the proposed angle-difference feedback framework in Sec. \ref{Sec: AD} without incorporating the CSI refinement framework in Sec. \ref{Sec:Refine}.
    Our Initial feedback model is trained with the mean squared error (MSE) distortion measure as $d(\mathbf{X},\mathbf{\hat X})=\Vert \mathbf{X} - \mathbf{\hat X}\Vert^2_{\rm F}$.
    In Fig.~\ref{fig: V_vs_angle}, we compare the NMSE performance of the CSI feedback frameworks with different quantization techniques and input forms:
    (i) Method in \cite{LB-SciFi},
    (ii) Method in \cite{EFNet},
    (iii) DL-based feedback with VQ (separate $\{\phi_{i,j}\}$, $\{\psi_{l,i}\}$),
    (iv) DL-based feedback with VQ ($\mathbf{\overline{V}}$),
    and (v) the proposed angle-difference feedback ($\mathbf{\Phi}$) with the unified VQ method in Sec. \ref{subSec: Time sparisty VQ}.
    For both (i) and (ii), uniform scalar quantization is applied to the latent vector; (i) first applies hyperparameterized preprocessing and separate compression to the angle parameters $\{\phi_{i,j}\}$ and $\{\psi_{l,i}\}$, while (ii) directly uses the beamforming matrix $\mathbf{\overline{V}}$ as the network input.
    Furthermore, we also consider feedback methods described in the IEEE 802.11 standard protocol for performance comparison. The standard feedback method considered in Fig. \ref{fig: V_vs_angle} applies uniform quantization to ($\phi$,$\psi$) with $(b_{\phi},b_{\psi})=(5,3)$ and $(6,4)$. The feedback overhead required by the IEEE 802.11 standard method is 6656 or 8320 bits for the DeepMIMO dataset, and 720 or 900 bits for the Wi-MIR dataset.
     Fig. \ref{fig: V_vs_angle} demonstrates that the proposed angle-difference feedback framework consistently outperforms the conventional methods, highlighting the effectiveness of the DL-based CSI feedback with the proposed framework and input design. Moreover, when combined with the proposed refinement framework, the angle-difference feedback method achieves superior performance compared to when the refinement is not applied.
     This result demonstrates the advantages of exploiting temporal correlation in the CSIs.

   \begin{figure}[t]
        \centering 
        {\epsfig{file=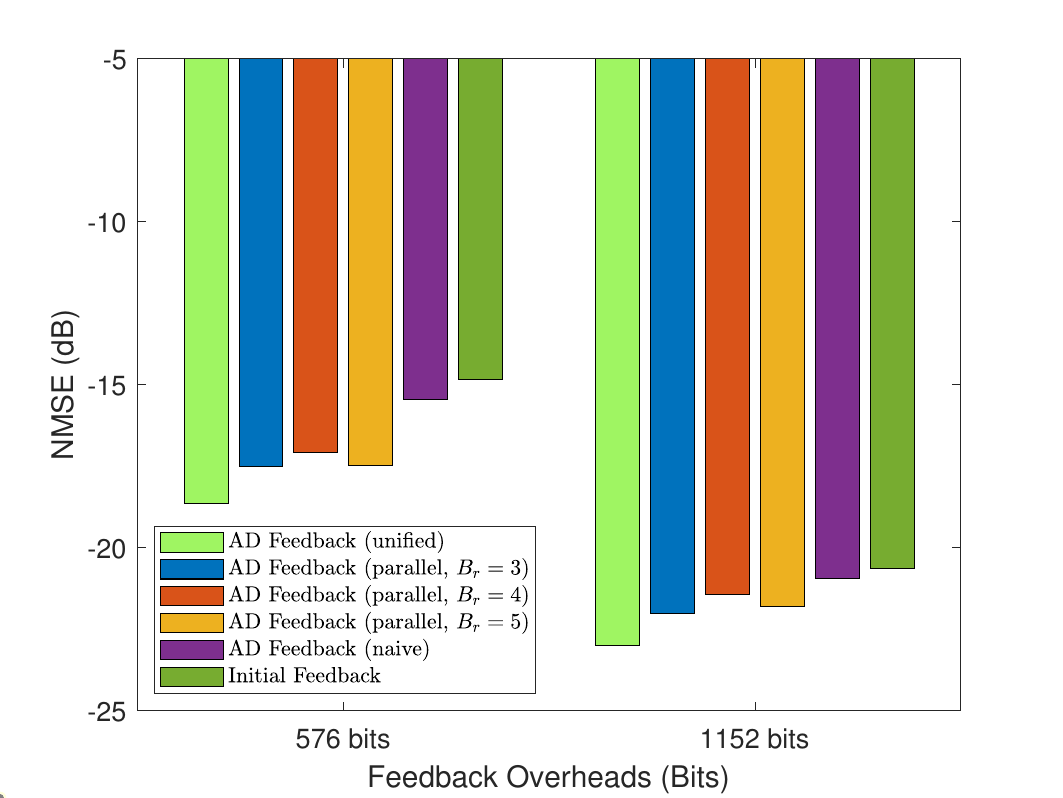, width=7.5cm}}
        \caption{Comparison of the NMSE (dB) performance of the angle-difference feedback framework with different VQ methods using the DeepMIMO dataset.}
        \label{fig: bar_graph_seqVQ}\vspace{-3mm}
    \end{figure}

    In Fig. \ref{fig: bar_graph_seqVQ}, we demonstrate the superiority of the DL-based angle-difference feedback framework presented in Sec. \ref{subSec: Multi-Stage VQ},E. 
    To this end, in Fig. \ref{fig: bar_graph_seqVQ}, we consider three different VQ frameworks: (i) {\em AD Feedback (naive)}, a naive angle-difference feedback framework that only utilizes the angle difference as an input without modifying the VQ module, (ii) {\em AD Feedback (parallel)}, the proposed angle-difference framework with the parallel VQ method in Sec. \ref{subSec: Multi-Stage VQ}, and (iii) {\em AD Feedback (unified)},  the proposed angle-difference framework with the unified VQ method in Sec. \ref{subSec: Time sparisty VQ}.  
    Fig. \ref{fig: bar_graph_seqVQ} illustrates the performance gains achieved by the three angle-difference frameworks compared to the original DL-based CSI feedback framework using the DeepMIMO dataset. These gains show the benefits of exploiting the temporal correlation of the CSIs. Furthermore, Fig. \ref{fig: bar_graph_seqVQ} demonstrates that the proposed frameworks outperform the naive angle-difference framework, emphasizing their effectiveness in mitigating error propagation inherent in the angle-difference feedback approach. Notably, the unified VQ method introduced in Sec. \ref{subSec: Time sparisty VQ} achieves a higher performance gain than the parallel VQ method discussed in Sec. \ref{subSec: Multi-Stage VQ}. This result demonstrates the advantages of the unified design of the VQ module and decoder, which effectively reduces both error propagation and reconstruction errors, while reducing the computational complexity of the feedback process.

    \begin{table}[t]
    \centering
    \caption{Comparison of the NMSE (\textit{dB}) gain of the proposed angle-difference feedback framework with various speed of STAs and bandwidths for the DeepMIMO dataset.}\label{table: correlation_comparison}
    {%
    \begin{tabular}{cccc}
    \hline
    \multicolumn{1}{c|}{\textbf{Speed of STAs}}                                       & \multicolumn{1}{c|}{\begin{tabular}[c]{@{}c@{}}Feedback\\ Overheads\end{tabular}} & \multicolumn{1}{c|}{Initial Feedback} & AD Feedback          \\ \hline
    \multicolumn{1}{c|}{\begin{tabular}[c]{@{}c@{}}0.4 m/s\\ or 0.56m/s\end{tabular}} & \multicolumn{1}{c|}{576 bits}                                                     & \multicolumn{1}{c|}{-14.85}           & -18.65               \\ \hline
    \multicolumn{1}{c|}{\begin{tabular}[c]{@{}c@{}}0.8 m/s\\ or 1.12m/s\end{tabular}} & \multicolumn{1}{c|}{576 bits}                                                     & \multicolumn{1}{c|}{-14.85}           & -17.22               \\ \hline
    \multicolumn{1}{c|}{\begin{tabular}[c]{@{}c@{}}2 m/s\\ or 2.8m/s\end{tabular}}    & \multicolumn{1}{c|}{576 bits}                                                     & \multicolumn{1}{c|}{-14.85}           & -17.39               \\ \hline
    \multicolumn{1}{l}{}                                                              & \multicolumn{1}{l}{}                                                              & \multicolumn{1}{l}{}                  & \multicolumn{1}{l}{} \\
    \multicolumn{1}{l}{}                                                              & \multicolumn{1}{l}{}                                                              & \multicolumn{1}{l}{}                  & \multicolumn{1}{l}{}
    \end{tabular}
    }
    {%
    \begin{tabular}{c|c|c|c}
    \hline
    \begin{tabular}[c]{@{}c@{}}\textbf{Bandwidth}\\ {\scriptsize{ (\# of total/sampled}}\\ {\scriptsize{ subcarriers)}}\end{tabular} & \begin{tabular}[c]{@{}c@{}}Feedback\\ Overheads\end{tabular} & Initial Feedback & AD Feedback \\ \hline
    \begin{tabular}[c]{@{}c@{}}20 MHz\\ (256/64)\end{tabular}                                        & 576 bits                                                     & -14.85           & -18.65      \\ \hline
    \begin{tabular}[c]{@{}c@{}}80 MHz\\ (1024/256)\end{tabular}                                      & 2592 bits                                                    & -17.80           & -20.35      \\ \hline
    \begin{tabular}[c]{@{}c@{}}320 MHz\\ (4096/1024)\end{tabular}                                    & 9504 bits                                                    & -17.31           & -20.09      \\ \hline
    \end{tabular}
    }
    \end{table}

    We also demonstrate the versatility of the proposed angle-difference feedback framework by varying the levels of temporal correlations and bandwidth settings in the dataset. In particular, we first consider three different levels of temporal correlation by setting the speed of STAs in the DeepMIMO dataset as (i) 0.4 m/s or 0.56 m/s, (ii) 0.8 m/s or 1.12 m/s, and (iii) 2 m/s or 2.8 m/s.
    Additionally, to analyze the impact of bandwidth setting, we vary the system bandwidth in the DeepMIMO dataset as (i) 20 MHz, (ii) 80 MHz, and (iii) 320 MHz. As shown in Table~\ref{table: correlation_comparison}, for each bandwidth setting, we indicate both the total number of subcarriers and the number of sampled subcarriers used for CSI feedback, denoted as {\em total/sampled subcarriers}. For instance, a 20 MHz system uses 256 total subcarriers from which 64 are sampled, while 80 MHz and 320 MHz systems use 1024/256 and 4096/1024, respectively.
    Table~\ref{table: correlation_comparison} demonstrates that the proposed angle-difference feedback framework consistently improves performance compared to using the initial feedback framework alone, across various levels of temporal correlations and bandwidth settings.
    
   \begin{figure}[t]
        \centering 
        \subfigure[DeepMIMO]{\epsfig{file=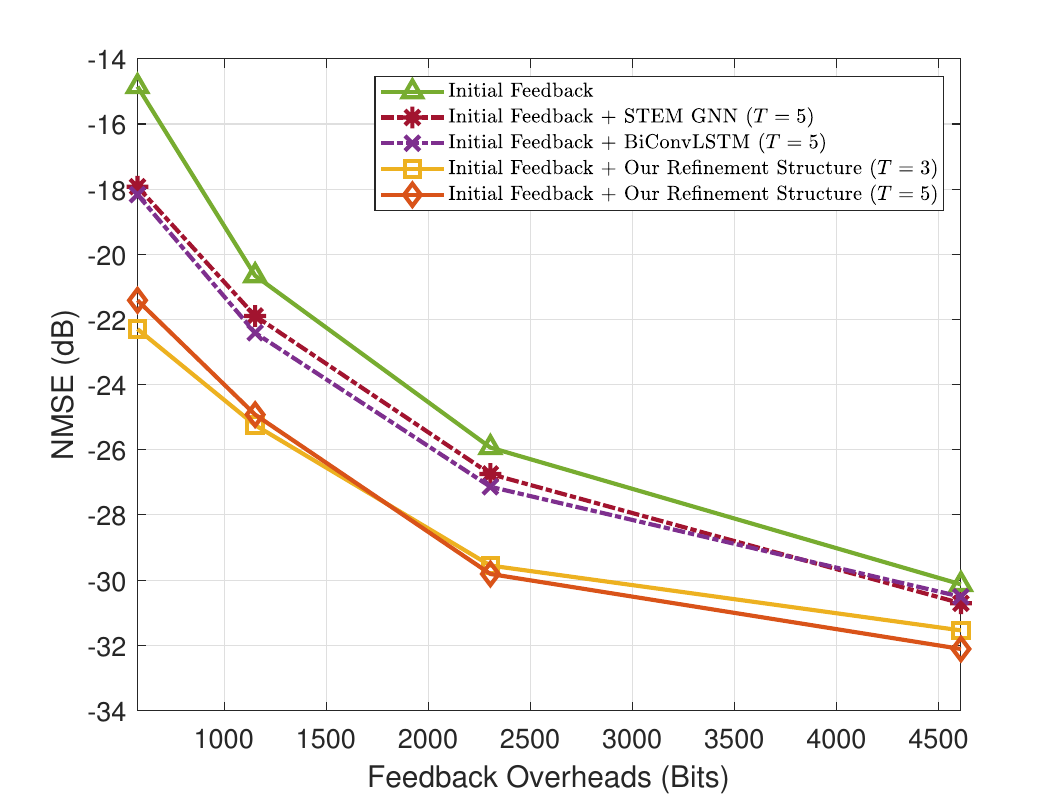, width=7.5cm}}
        
        \subfigure[Wi-MIR]{\epsfig{file=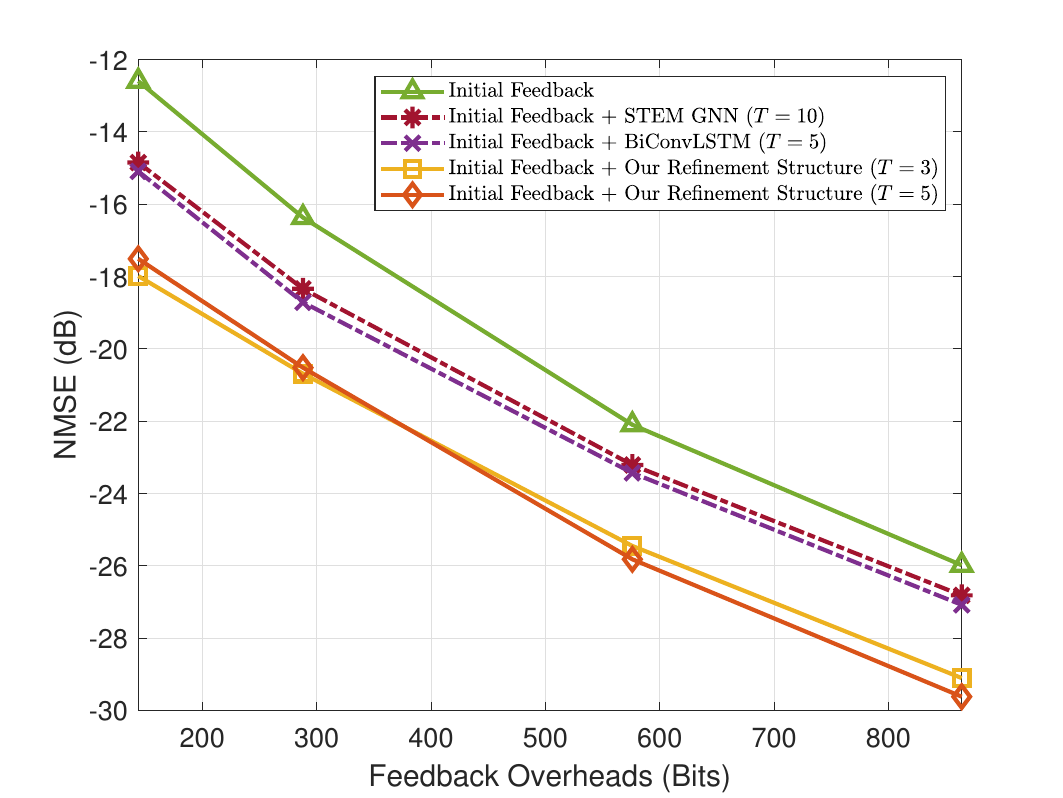, width=7.5cm}}
        \caption{Comparison of the NMSE (dB) performance of various CSI refinement structures when integrated with the proposed CSI feedback framework.}
        \label{fig: performance_nmse_refine_comparison}\vspace{-3mm}
    \end{figure}

    \begin{table*}[t]
    \centering
    \caption{Comparison of the memory and computational efficiency of DL-based CSI feedback and Refinement models.}\label{table: complexity}
    \resizebox{\textwidth}{!}{
    \begin{tabular}{cccccccccccccc}
    \hline
    \multicolumn{14}{c}{\textbf{DeepMIMO Dataset}}                                                                                                                                                                                                                                                                                                                                                                                                     \\ \hline
    \multicolumn{7}{c||}{\textbf{CSI Feedback Module}}                                                                                                                                                                                  & \multicolumn{7}{c}{\textbf{CSI Refinement Module}}                                                                                                                                                            \\ \hline
    \multicolumn{1}{c|}{\multirow{2}{*}{Feedback Overheads}} & \multicolumn{2}{c|}{Parameters (M)}              & \multicolumn{2}{c|}{FLOPs (M)}                    & \multicolumn{2}{c||}{Processing Time (s)}                         & \multicolumn{1}{c|}{\multirow{2}{*}{Model Type, $T$}} & \multicolumn{2}{c|}{Parameters (M)}              & \multicolumn{2}{c|}{FLOPs (M)}                       & \multicolumn{2}{c}{Processing Time (s)}     \\ \cline{2-7} \cline{9-14} 
    \multicolumn{1}{c|}{}                                    & STA                  & \multicolumn{1}{c|}{AP}   & STA                  & \multicolumn{1}{c|}{AP}    & STA                  & \multicolumn{1}{c||}{AP}                   & \multicolumn{1}{c|}{}                                 & STA                  & \multicolumn{1}{c|}{AP}   & STA                  & \multicolumn{1}{c|}{AP}       & STA                  & AP                   \\ \hline
    \multicolumn{1}{c|}{\textbf{576 bits}}                   & 0.22                 & \multicolumn{1}{c|}{0.42} & 10.59                & \multicolumn{1}{c|}{27.56} & 4.41$\times 10^{-4}$ & \multicolumn{1}{c||}{5.80$\times 10^{-4}$} & \multicolumn{1}{c|}{\textbf{Our Structure, $T=3$}}    & -                    & \multicolumn{1}{c|}{2.95} & -                    & \multicolumn{1}{c|}{202.13}   & -                    & 1.23$\times 10^{-2}$ \\
    \multicolumn{1}{c|}{\textbf{1152 bits}}                  & 0.28                 & \multicolumn{1}{c|}{0.50} & 12.51                & \multicolumn{1}{c|}{28.89} & 4.66$\times 10^{-4}$ & \multicolumn{1}{c||}{6.01$\times 10^{-4}$} & \multicolumn{1}{c|}{\textbf{Our Structure, $T=5$}}    & -                    & \multicolumn{1}{c|}{2.96} & -                    & \multicolumn{1}{c|}{209.28}   & -                    & 1.44$\times 10^{-2}$ \\
    \multicolumn{1}{c|}{\textbf{2304 bits}}                  & 0.44                 & \multicolumn{1}{c|}{0.64} & 16.34                & \multicolumn{1}{c|}{31.54} & 5.31$\times 10^{-4}$ & \multicolumn{1}{c||}{6.48$\times 10^{-4}$} & \multicolumn{1}{c|}{STEM GNN, $T=5$}                  & -                    & \multicolumn{1}{c|}{8.53} & -                    & \multicolumn{1}{c|}{18851.13} & -                    & 4.24$\times 10^{-1}$ \\
    \multicolumn{1}{c|}{\textbf{4608 bits}}                  & 0.73                 & \multicolumn{1}{c|}{0.94} & 24.01                & \multicolumn{1}{c|}{36.85} & 6.51$\times 10^{-4}$ & \multicolumn{1}{c||}{7.91$\times 10^{-4}$} & \multicolumn{1}{c|}{BiConvLSTM, $T=5$}                & -                    & \multicolumn{1}{c|}{0.30} & -                    & \multicolumn{1}{c|}{1281.14}  & -                    & 8.18$\times 10^{-2}$ \\ \hline
    \multicolumn{1}{l}{}                                     & \multicolumn{1}{l}{} & \multicolumn{1}{l}{}      & \multicolumn{1}{l}{} & \multicolumn{1}{l}{}       & \multicolumn{1}{l}{} & \multicolumn{1}{l}{}                      & \multicolumn{1}{l}{}                                  & \multicolumn{1}{l}{} & \multicolumn{1}{l}{}      & \multicolumn{1}{l}{} & \multicolumn{1}{l}{}          & \multicolumn{1}{l}{} & \multicolumn{1}{l}{} \\
    \multicolumn{1}{l}{}                                     & \multicolumn{1}{l}{} & \multicolumn{1}{l}{}      & \multicolumn{1}{l}{} & \multicolumn{1}{l}{}       & \multicolumn{1}{l}{} & \multicolumn{1}{l}{}                      & \multicolumn{1}{l}{}                                  & \multicolumn{1}{l}{} & \multicolumn{1}{l}{}      & \multicolumn{1}{l}{} & \multicolumn{1}{l}{}          & \multicolumn{1}{l}{} & \multicolumn{1}{l}{} \\
    \multicolumn{1}{l}{}                                     & \multicolumn{1}{l}{} & \multicolumn{1}{l}{}      & \multicolumn{1}{l}{} & \multicolumn{1}{l}{}       & \multicolumn{1}{l}{} & \multicolumn{1}{l}{}                      & \multicolumn{1}{l}{}                                  & \multicolumn{1}{l}{} & \multicolumn{1}{l}{}      & \multicolumn{1}{l}{} & \multicolumn{1}{l}{}          & \multicolumn{1}{l}{} & \multicolumn{1}{l}{} \\ \hline
    \multicolumn{14}{c}{\textbf{Wi-MIR Dataset}}                                                                                                                                                                                                                                                                                                                                                                                                       \\ \hline
    \multicolumn{7}{c||}{\textbf{CSI Feedback Module}}                                                                                                                                                                                  & \multicolumn{7}{c}{\textbf{CSI Refinement Module}}                                                                                                                                                            \\ \hline
    \multicolumn{1}{c|}{\multirow{2}{*}{Feedback Overheads}} & \multicolumn{2}{c|}{Parameters (M)}              & \multicolumn{2}{c|}{FLOPs (M)}                    & \multicolumn{2}{c||}{Processing Time (s)}                         & \multicolumn{1}{c|}{\multirow{2}{*}{Model Type, $T$}} & \multicolumn{2}{c|}{Parameters (M)}              & \multicolumn{2}{c|}{FLOPs (M)}                       & \multicolumn{2}{c}{Processing Time (s)}     \\ \cline{2-7} \cline{9-14} 
    \multicolumn{1}{c|}{}                                    & STA                  & \multicolumn{1}{c|}{AP}   & STA                  & \multicolumn{1}{c|}{AP}    & STA                  & \multicolumn{1}{c||}{AP}                   & \multicolumn{1}{c|}{}                                 & STA                  & \multicolumn{1}{c|}{AP}   & STA                  & \multicolumn{1}{c|}{AP}       & STA                  & AP                   \\ \hline
    \multicolumn{1}{c|}{\textbf{144 bits}}                   & 0.16                 & \multicolumn{1}{c|}{0.36} & 3.51                 & \multicolumn{1}{c|}{8.83}  & 1.51$\times 10^{-4}$ & \multicolumn{1}{c||}{2.19$\times 10^{-4}$} & \multicolumn{1}{c|}{\textbf{Our Structure, $T=3$}}    & -                    & \multicolumn{1}{c|}{2.95} & -                    & \multicolumn{1}{c|}{49.93}    & -                    & 5.61$\times 10^{-3}$ \\
    \multicolumn{1}{c|}{\textbf{288 bits}}                   & 0.18                 & \multicolumn{1}{c|}{0.38} & 3.99                 & \multicolumn{1}{c|}{9.16}  & 1.71$\times 10^{-4}$ & \multicolumn{1}{c||}{2.35$\times 10^{-4}$} & \multicolumn{1}{c|}{\textbf{Our Structure, $T=5$}}    & -                    & \multicolumn{1}{c|}{2.96} & -                    & \multicolumn{1}{c|}{51.41}    & -                    & 5.75$\times 10^{-3}$ \\
    \multicolumn{1}{c|}{\textbf{576 bits}}                   & 0.21                 & \multicolumn{1}{c|}{0.42} & 4.94                 & \multicolumn{1}{c|}{9.82}  & 2.21$\times 10^{-4}$ & \multicolumn{1}{c||}{2.69$\times 10^{-4}$} & \multicolumn{1}{c|}{STEM GNN, $T=10$}                 & -                    & \multicolumn{1}{c|}{2.91} & -                    & \multicolumn{1}{c|}{529.53}   & -                    & 2.92$\times 10^{-2}$ \\
    \multicolumn{1}{c|}{\textbf{864 bits}}                   & 0.25                 & \multicolumn{1}{c|}{0.46} & 5.90                 & \multicolumn{1}{c|}{10.49} & 2.61$\times 10^{-4}$ & \multicolumn{1}{c||}{2.98$\times 10^{-4}$} & \multicolumn{1}{c|}{BiConvLSTM, $T=5$}                & -                    & \multicolumn{1}{c|}{0.30} & -                    & \multicolumn{1}{c|}{272.79}   & -                    & 1.82$\times 10^{-2}$ \\ \hline
    \end{tabular}}
    \end{table*}

   \subsection{Performance of the Proposed CSI Refinement Framework}\label{subSec: refine_performance}
    In this subsection, we demonstrate the superiority of the proposed CSI refinement framework in Sec. \ref{Sec:Refine}. 
    In Fig. \ref{fig: performance_nmse_refine_comparison}, we evaluate the NMSE performance of various CSI refinement structures when integrated with the proposed CSI feedback framework in Sec. \ref{Sec: VQ-VAE}. To isolate the impact of the CSI refinement module, the simulations do not include the angle-difference feedback strategy. For comparison, we consider two existing DL-based CSI refinement methods. The first is the spectral temporal graph neural network (STEM GNN) from \cite{STEM}, which utilizes graph neural networks and spectral analysis to capture temporal correlation, as applied in \cite{STEM_CSI}. The second is the bi-directional convolutional LSTM (BiConvLSTM), adopted in \cite{BiLSTM_CSI1, BiLSTM_CSI2, BiLSTM_CSI3}, which enhances the standard LSTM structure by integrating convolutional layers and enabling both forward and backward data processing. For a fair comparison, each method is evaluated with an optimized time window size $T$ to maximize performance. Table~\ref{table: complexity} summarizes the number of parameters (in millions) and computational complexities for all methods and datasets.
    The computational complexity is reported in both floating-point operations (FLOPs), measured in millions, and processing time, measured in seconds. Here, FLOPs refer to the total number of multiply-add operations required for a single forward pass through the model. The values required at the STA and AP are reported separately, across the CSI feedback and CSI refinement modules. All model parameters are represented in 32-bit floating-point format (4 bytes), so the memory usage (in bytes) can be computed by multiplying the number of parameters by 4. The processing time is measured as the average forward pass time on a system equipped with an NVIDIA GeForce RTX 4090 GPU and an AMD Ryzen 9 9950X CPU.
    Fig. \ref{fig: performance_nmse_refine_comparison} demonstrates that the proposed CSI refinement framework, leveraging our refinement structure, provides substantial NMSE improvement over the baseline without CSI refinement. Moreover, the results from Fig. \ref{fig: performance_nmse_refine_comparison} and Table~\ref{table: complexity} confirm that our refinement structure outperforms both existing methods in terms of NMSE while maintaining lower computational complexity.

    
    \subsection{Throughput Performance}

    We consider throughput as another key performance metric in our simulations. To assess the impact on system-level throughput, we conduct simulations in MATLAB R2024a environment. In the current IEEE 802.11 standards, the system determines the data rate considering how accurately the symbols are received based on the error vector magnitude (EVM) defined as
    \begin{align}\label{eq: EVM}
    \text{EVM} = \frac{\mathbb{E}\left[\Vert\mathbf{s}[k]-\mathbf{\hat s}[k]\Vert^2\right]}{\mathbb{E}\left[\Vert\mathbf{s}[k]\Vert^2\right]},
    \end{align}
    where $\mathbf{s}[k]$ and $\mathbf{\hat s}[k]$ are transmitted symbols and received symbols, respectively. Based on the value of EVM, the system adaptively determines the modulation type and coding rate as specified in Table~\ref{table: EVM}. The EVM coefficient $\gamma$ in Table~\ref{table: EVM} represents relative data rate.
    The gross throughput is determined directly from the value of $\gamma$, which is the data rate achieved without considering CSI feedback process. The gross throughput, denoted as $r$, is computed as
    \begin{align}\label{eq: gross throughput}
        r = \frac{N_{\rm sp}}{N_{\rm fft}+N_{\rm cp}}\times N_s\times b \times \gamma,
    \end{align}
    where $N_{\rm fft}$ is the number of FFT points, $N_{\rm cp}$ is the length of cyclic prefix,  $N_{\rm sp}$ is the number of subcarriers carrying payload, and $b$ is the sampling rate. In the our simulations for the throughput analysis, we set $N_{\rm fft} =256$, $N_{\rm cp}=32$, $N_{\rm sp}=234$, and $b=20$MHz. Unlike the gross throughput, the {\em net} throughput stands for the data rate where the CSI feedback overhead is considered. The net throughput, denoted as $\bar{r}$, is computed as
    \begin{align}\label{eq: net throughput}
    \bar{r} = \frac{N_{\rm data}}{N_{\rm data}+N_{\rm overhead}}\times r,
    \end{align}
    where $N_{\rm data}$ is the duration for data transmission and ${N}_{\rm overhead}$ is the total duration for transmitting NDPA, NDP, ACK, and CBR. For the CSI feedback, which is corresponded to CBR, the BPSK and coding rate of 1/2 are employed. Also, the length of data packet is set to 2000 bytes.
    
    In Table~\ref{table: throughput}, we compare the throughput performance of various CSI feedback methods in Wi-Fi systems using the DeepMIMO dataset. For the IEEE 802.11 standard feedback, T0 and T1 feedback apply uniform quantization to ($\phi$,$\psi$) with $(b_{\phi},b_{\psi})=(7,5)$ and $(9,7)$, respectively. Table~\ref{table: throughput} shows that the proposed frameworks, both with and without the angle-difference feedback and refinement module, outperform the standard CSI feedback methods. Additionally, the proposed framework, when integrated with the refinement module, achieves superior net throughput performance compared to its counterpart without these enhancements. 

     \begin{table*}[t]
    \centering
    \caption{\scriptsize EVM coefficient $\gamma$ specified in IEEE 802.11\textit{be} standard\cite{Wi-Fi7}.}\label{table: EVM}
    {\scriptsize
    \begin{tabular}{c||c|c|c|c|c|c|c|c|c}
    \hline
    \textbf{EVM (dB)}  & {[}-10, -13) & {[}-13, -16) & {[}-16, -19) & {[}-19, -22) & {[}-22, -25) & {[}-25, -27) & {[}-27, -30) & {[}-30, -32) & {[}-32, -35) \\ \hline
    \textbf{Modulation}  & QPSK & QPSK & 16-QAM & 16-QAM & 64-QAM & 64-QAM & 64-QAM & 256-QAM & 256-QAM \\ \hline
    \textbf{Coding Rate}  & 1/2 & 3/4 & 1/2 & 3/4 & 2/3 & 3/4 & 5/6 & 3/4 & 5/6 \\ \hline
    \textbf{$\gamma$}  & 1 & 1.5 & 2 & 3 & 4 & 4.5 & 5 & 6 & 20/3 \\ \hline
    \end{tabular}}
    \end{table*}

    \begin{table*}[t]
    \centering
    \caption{Comparison of throughput performance of various combinations of the proposed frameworks using the DeepMIMO dataset.}\label{table: throughput}
    {
    \begin{tabular}{c|cccc||c|cccc}
    \hline
    \begin{tabular}[c]{@{}c@{}}Feedback\\ Scheme\end{tabular} & \begin{tabular}[c]{@{}c@{}}Feedback\\ Overhead\\ (Bits)\end{tabular} & EVM (dB) & \begin{tabular}[c]{@{}c@{}}Gross\\ Throughput\\ (Mb/s)\end{tabular} & \begin{tabular}[c]{@{}c@{}}Net\\ Throughput\\ (Mb/s)\end{tabular} & \begin{tabular}[c]{@{}c@{}}Feedback\\ Scheme\end{tabular} & \begin{tabular}[c]{@{}c@{}}Feedback\\ Overhead\\ (Bits)\end{tabular} & EVM (dB) & \begin{tabular}[c]{@{}c@{}}Gross\\ Throughput\\ (Mb/s)\end{tabular} & \begin{tabular}[c]{@{}c@{}}Net\\ Throughput\\ (Mb/s)\end{tabular} \\ \hline
    Standard (T0) & 9984 & -28.57 & 162.5 & 19.37 & Standard (T1) & 13312 & -29.28 & 162.5 & 15.52 \\ \hline
    \multirow{4}{*}{\textbf{Initial Feedback}} & 576 & -21.88 & 97.5 & 51.15 & \multirow{4}{*}{\textbf{\begin{tabular}[c]{@{}c@{}} AD Feedback\end{tabular}}} & 577 & -24.74 & 130 & \textbf{58.88} \\
     & 1152 & -25.92 & 146.25 & \textbf{54.51} &  & 1153 & -26.91 & 146.25 & 54.51 \\
     & 2304 & -28.04 & 162.5 & 45.26 &  & 2305 & -28.24 & 162.5 & 45.26 \\
     & 4608 & -28.82 & 162.5 & 32.31 &  & 4609 & -28.95 & 162.5 & 32.31 \\ \hline
    \multirow{4}{*}{\textbf{\begin{tabular}[c]{@{}c@{}}Initial Feedback\\ + Refinement\end{tabular}}} & 576 & -26.72 & 146.25 & \textbf{62.00} &  \multirow{4}{*}{\textbf{\begin{tabular}[c]{@{}c@{}}AD Feedback\\ + Refinement\end{tabular}}}  & 577 & -27.22 & 162.5 & \textbf{64.74} \\
     & 1152 & -27.68 & 162.5 & 56.62 &  & 1153 & -28.15 & 162.5 & 56.62 \\
     & 2304 & -28.78 & 162.5 & 45.26 &  & 2305 & -28.72 & 162.5 & 45.26 \\
     & 4608 & -28.89 & 162.5 & 32.31 &  & 4609 & -29.10 & 162.5 & 32.31 \\ \hline
    \end{tabular}}  
    \end{table*}

    \section{Conclusion}
    In this paper, we have presented a DL-based CSI feedback framework for Wi-Fi systems. Our framework employs the VQ approach to enable efficient finite-bit representation of the latent vector within an autoencoder architecture. We have proposed using an angle parameter matrix as the input for the encoder network and developed angle-difference feedback and  CSI refinement module to enhance feedback efficiency by leveraging the temporal correlation of CSI. Simulation results demonstrate that the proposed frameworks outperform the  IEEE 802.11 standard protocols. Additionally, our results show that utilizing CSI's temporal correlation leads to significant performance gains.

    An important direction for future research is to extend our strategy for a multi-user MIMO system model, in order to serve multi users effectively in the Wi-Fi systems. Another important research direction is to develop an advanced generalization approach for our strategy, in order to provide the robustness to dynamic and unpredictable changes in channel distributions. It would also be possible to improve the performance of our strategy by applying an entropy coding to the VQ output.

\end{document}